\documentclass[lettersize,journal]{IEEEtran}
\usepackage{amsmath,amsfonts}
\usepackage{array}
\usepackage[caption=false,font=normalsize,labelfont=sf,textfont=sf]{subfig}
\usepackage{textcomp}
\usepackage{stfloats}
\usepackage{url}
\usepackage{verbatim}
\usepackage{graphicx}
\usepackage{cite}
\usepackage{caption}
\usepackage{booktabs}  
\usepackage[lined,boxed,commentsnumbered,ruled]{algorithm2e}
\usepackage{subfig}
\newcommand\figref{Fig.~\ref}
\newcommand\algref{Alg.~\ref}
\newcommand\tabref{Tab.~\ref}
\newcommand\secref{Section~\ref}

\begin{document}

\title{Efficient Gridless DoA Estimation Method of Non-uniform Linear Arrays with Applications in Automotive Radars}

\author{Silin Gao, Zhe Zhang,~\IEEEmembership{Member,~IEEE}, Muhan Wang, Yan Zhang, Jie Zhao, Bingchen Zhang, Yue Wang,~\IEEEmembership{Senior Member, IEEE} and Yirong Wu
  \thanks{Corresponding author: Zhe Zhang (zhangzhe01@aircas.ac.cn)}
  \thanks{
    S. Gao, Z. Zhang, M. Wang, Y. Zhang, B. Zhang and Y. Wu are with Aerospace Information Research Institute, Chinese Academy of Sciences, Beijing, China.}
  \thanks{
    S. Gao, M. Wang, Y. Zhang, B. Zhang and Y. Wu are also with Key Laboratory of Technology in Geo-spatial Information Processing and Application System, Chinese Academy of Sciences, Beijing, China, and School of Electronic, Electrical and Communication Engineering, University of Chinese Academy of Sciences, Beijing, China.}
  \thanks{
    Z. Zhang and M. Wang are also with Suzhou Key Laboratory of Microwave Imaging, Processing and Application Technology, and Suzhou Aerospace Information Research Institute, Suzhou, Jiangsu, China.}
  \thanks{Jie Zhao is with the Beijing Autoroad Tech Co., Ltd.}
  \thanks{Yue Wang is with Electrical and Computer Engineering Department, George Mason University, Fairfax, VA, USA.}
}

\maketitle

\begin{abstract}
  This paper focuses on the gridless direction-of-arrival (DoA) estimation for data acquired by non-uniform linear arrays (NLAs) in automotive applications. Atomic norm minimization (ANM) is a promising gridless sparse recovery algorithm under the Toeplitz model and solved by convex relaxation, thus it is only applicable to uniform linear arrays (ULAs) with array manifolds having a Vandermonde structure.
  In automotive applications, it is essential to apply the gridless DoA estimation to NLAs with arbitrary geometry with efficiency.
  In this paper, a fast ANM-based gridless DoA estimation algorithm for NLAs is proposed, which employs the array manifold separation technique and the accelerated proximal gradient (APG) technique, making it applicable to NLAs without losing of efficiency. Simulation and measurement experiments on automotive multiple-input multiple-outpt (MIMO) radars demonstrate the superiority of the proposed method.

\end{abstract}


\section{Introduction}
\IEEEPARstart{D}{irection}-of-arrival (DoA) estimation is an important area in array signal processing, which holds paramount importance in a multitude of applications, including but not limited to radar, sonar, wireless communication, and microphone arrays. DoA estimation is of critical importance in automotive radar systems, which are used in advanced driver assistance systems (ADAS) and autonomous vehicles for safety and navigation purposes. Automotive radar systems rely on DoA estimation to accurately detect and locate objects such as other vehicles, pedestrians, and obstacles on the road.

Plenty of DOA algorithms have been proposed in the last decades. Classical beamforming methods, such as Bartlett algorithm and minimum variance distortionless response (MVDR) algorithm \cite{mvdr} can effectively perform filtering in the spatial domain thus enabling DoA estimation. However, the theoretical resolution depends mainly on the array aperture and strictly limited by the Rayleigh resolution.
Subspace algorithms, such as the multiple signal classification (MUSIC) \cite{music,rootmusic} algorithm and the estimation of signal parameters via rotational invariance techniques (ESPRIT) \cite{esprit,esprit1995,esprit2014,esprit2020} algorithm has te super-resolution ability. However, it needs multiple snapshots (multiple measurement vector or MMV) of data and a high signal-to-noise ratio (SNR), as well as incoherent signal sources.

In the recent decades, sparse signal processing based techniques such as compressed sensing (CS) techniques \cite{CSDonoho,CSBaraniuk,CSCandes} has been introduced into DoA estimation by exploiting the sparsity structure of the signal. \cite{l1svd,l1sracv} solves the DoA estimation problem by leveraging the spatial domain discretization and the $\ell _1$ norm regularization model, making it applicable to single snapshot (SMV) and/or coherent scenarios. However, the spatial domain discretization suffers the grid mismatching problem\cite{off}. Tackling this issue, the straightforward approach is to increase the grid density, such as the iterative grid refinement (IGR) technique \cite{igr}. However, using a dense set of grids will violate the restricted isometry property (RIP)  \cite{rip} thus resulting the sparse reconstruction impossible.

Aiming at the gridding effect, the off-grid DOA estimation model is first proposed in \cite{S-TLS} to parameterize the off-grid error as a model perturbation. References \cite{bayesian,bayesian1,bayesian2,bayesian3} improve the off-grid signal model and solve it via sparse Bayesian learning (SBL), but suffering from complicated parameter tuning and high computational load.

Atomic norm minimization (ANM) is the cutting-edge development of sparse signal processing \cite{TANGANM}. Different from the perturbation based methods, ANM directly constructed a gridless sparse model from scratch. Retaining the benefit of gridless output and super-resolution ability, ANM does not require incoherent signal sources and works smoothly with single snapshot. However, currently ANM is mainly solved via semi-positive definite programming (SDP) and relies on the Vandermonde structure of atoms. For array signal processing, only data acquired by uniform linear arrays (ULAs) can be processed by the ANM algorithm. \cite{array1,array2,array3,array4,array5,incomplete} generalize ANM to the sparse arrays or incomplete data. However, in practical situations, it is often not possible to select the array geometry from some narrow categories. Therefore, it is of great significance to study the ANM algorithm that applicable to arbitrary non-uniform linear arrays (NLAs).

In this paper, we study the gridless DoA estimation problem for off-grid targets in automotive radar applications. We propose a fast gridless DoA estimation algorithm for NLAs based on ANM, namely FNLANM algorithm. The proposed algorithm uses the array manifold separation technique \cite{manifold1,manifold2,manifold3} to transform the physical (non-uniform) array into a virtual uniform array. The original array manifold is given as the product of a sampling matrix and a (uniform) Vandermonde vector, thus expanding the capability of ANM algorithm to arbitrary linear arrays. This optimization problem can also be solved via SDP, but the computational complexity would be impracticable for large virtual arrays. Inspired by \cite{ivdst}, we introduce the accelerated proximal gradient (APG) technique into the proposed algorithm to replace the SDP, reducing the computational complexity from $\mathcal{O}(N_v^{3.5})$ to $\mathcal{O}(N_v^{2})$, where $N_v$ is the number of virtual array elements.

The rest of this paper is organized as follows.
\secref{sec:Auto} describes the background of automotive radars, establishes the signal model, and introduces the grid-based CS algorithms for DoA estimation.
The concept of ANM and gridless DoA estimation algorithm for ULAs are given in \secref{sec:method}. The proposed FNLANM gridless DoA estimation algorithm is also introduced in this section.
Numerical analysis are discussed in \secref{sec:simu}, followed by the measured data results shown in \secref{sec:real}.
Finally, the conclusions are given in \secref{sec:con}.

Throughout this paper, vectors are denoted by lowercase bold characters (e.g. $\mathbf{v}$) and matrices by uppercase bold characters (e.g. $\mathbf{M})$. $(\cdot)^T$ is the transpose operator and $(\cdot)^H$ is the Hermitian transpose operator.

\section{Backgrounds}\label{sec:Auto}
This section provides some preliminary backgrounds of this paper, including automotive radar model, NLA signal model, and canonical compressed sensing based DoA estimation methods.

\subsection{Automotive radars}

Automotive radars transmit electromagnetic wave and receive echoes to detect nearby objects. To obtain both the range and velocity of targets, frequency modulated continuous wave (FMCW) is commonly used as the transmitting waveform in automotive radars, which can provide good detection performance in complex traffic situations with multiple targets \cite{review}.


Commercial FMCW radars, which usually work at 77 GHz mmWave band, can easily achieve high resolution in range and velocity.
To achieve angular resolving ability, automotive radars have evolved from using ULAs to multiple-input multiple-output (MIMO) arrays. ULAs that satisfying the Nyquist sampling theorem are the most commonly employed arrays. For ULAs, improving the angular resolution requires enlarged array aperture as well as huge amount of elements to, which implies increased hardware cost and computational complexity \cite{ula}. In recent years, MIMO technology has become increasingly popular in the design of automotive radars. This is primarily due to its ability to enhance radar performance and boost angular resolution, thereby improving the overall capabilities \cite{mimo1,mimo2}. Constructing virtual arrays with larger apertures is the technique employed in MIMO radars to achieve higher angular resolution, and the virtual arrays are usually NLAs.

NLAs offer many advantages over ULAs with the same number of elements. The non-uniform spacing between elements provides more information about about the received signal, which can improve the angular resolution. NLAs can have reduced sidelobe levels compared to ULAs, which can improve the ability to detect weak targets in the presence of noise and interference. NLAs offers more design flexibility and lower cost of second design and implementation than ULAs with the same performance.

Overall, the use of NLAs in automotive radar systems offers a number of benefits that can significantly enhance the overall performance. Therefore, it is necessary for us to study advanced signal processing methods applicable to NLAs, especially the DoA estimation algorithms.

\subsection{Signal model}

It is assumed that the linear array consists of $N$ array elements. With $E_0$ as the reference array element, the positions of each array elements are written as $\mathbf{r}=
  \begin{bmatrix}
    0 & r_1 & \cdots & r_{N-1}
  \end{bmatrix}^T$.
Note that we do not assume the $\mathbf{r}$ to be evenly distributed, i.e. it can be an arbitrary configured linear array.
In general, we assume that the array elements are isotropic, that there is no error in spatial location, no coupling among the array elements, and no extra error in the received response. The sources are assumed to be far-field, so the received signal is approximated as a plane wave when it arrives at the array. The subsequent sections are based on those assumptions.

\begin{figure}[htbp]
  \centering
  \includegraphics[width=\linewidth]{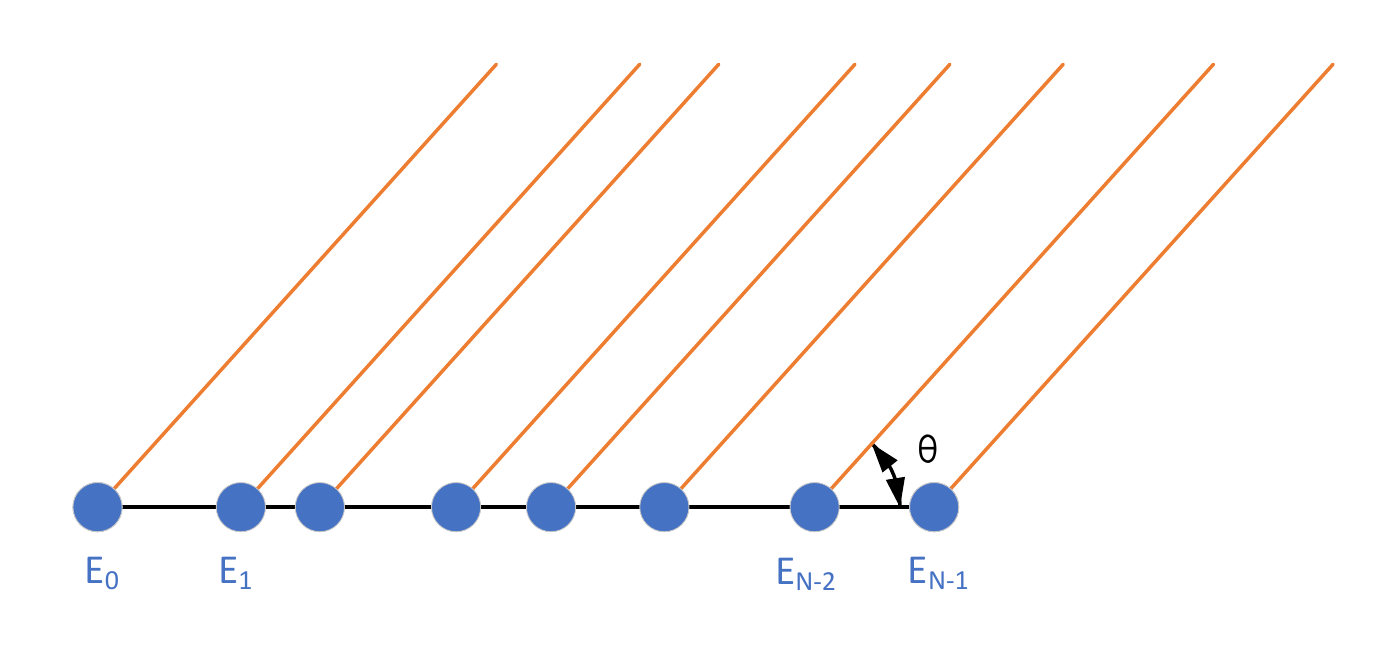}
  \caption{Linear array schematic.}
  \label{nla}
\end{figure}

\figref{nla} shows a schematic diagram of a linear array receiving signals with DoA angle $\theta$. The array manifold of this linear array can be expressed as:
\begin{equation}
  \mathbf{a}(\theta)=
  \begin{bmatrix}
    1 & e^{j2\pi\frac{r_1}{\lambda}\cos\theta} & \cdots & e^{j2\pi\frac{r_{N-1}}{\lambda}\cos\theta}
  \end{bmatrix}^T.
\end{equation}

For a noisy scenario with $K$ targets located at $K$ different directions, the signal $\mathbf{x}$ received by the array can be written as:

\begin{equation}
  \mathbf{x}=\sum_{k=1}^K c_k \mathbf{a}(\theta_k)+\mathbf{n},
\end{equation}
where \begin{itemize}
  \item $c_k$ is the reflectivity of the $k$-th target;
  \item $\theta_k$ is the direction of the $k$-th target;
  \item $\mathbf{n}$ is the additive Gaussian white noise.
\end{itemize}

We need to recover $\{(c_k, \theta_k)\}$ from the observed $\mathbf{x}$. There are a number of well-established algorithms available to accurately locate targets in linear arrays. Currently, the most widely used method for single snapshot NLA DoA estimation is compressed sensing, or CS based methods.

\subsection{DoA via CS method}

CS is a grid-based method, which discretize $\theta\in(0,\pi)$ into $M$ virtual grids and define a basis set of steering vectors $\mathbf{a}(\theta)$ as $\mathcal{A}_M=\{\mathbf{a}(\theta):\theta\in(0,\frac{1}{M}\pi,\cdots,\frac{M-1}{M}\pi)\}$.
Therefore, the sparsity of the scenario can be defined as the $\ell _0$ norm (and relaxed to the $\ell _1$ norm) of $\mathbf{x}$ over the basis $\mathcal{A}_M$ as:
\begin{equation}\label{csl0}
  \lVert \mathbf{x} \rVert _{\mathcal{A}_M,0}=\text{inf}\left\{ K:\,\,\sum_{k=1}^K{\hat{c}_k\mathbf{a}\left( \hat{\theta}_k \right) =\mathbf{x}},\,\,\mathbf{a}\left( \hat{\theta}_k \right) \in \mathcal{A}_M \right\} ,
\end{equation}
\begin{equation}\label{csl1}
  \lVert \mathbf{x} \rVert _{\mathcal{A}_M,1}=\text{inf}\left\{ \sum_{k=1}^K|\hat{c}_k|:\,\,\sum_{k=1}^K{\hat{c}_k\mathbf{a}\left( \hat{\theta}_k \right) =\mathbf{x}},\,\,\mathbf{a}\left( \hat{\theta}_k \right) \in \mathcal{A}_M \right\} .
\end{equation}

In the compressed sensing framework, it is proved that we can recover $\left\{ \theta_k \right\}$ and $\left\{ c_k \right\}$ via relaxed convex $\ell _1$ norm minimization under some conditions \cite{rip}. Therefore, the DoA information of targets can be obtained by solving the optimization problem based on $\ell _1$ norm minimization as follows:

\begin{equation}\label{cs}
  \min_\mathbf{c} \frac{1}{2}\lVert \mathbf{x}-\mathbf{Ac} \rVert_2^2 +\gamma\lVert \mathbf{x} \rVert_{\mathcal{A}_M,1},
\end{equation}
where \begin{itemize}
  \item $\mathbf{A}$ is the $N \times M$ observation matrix with $[\mathbf{A}]_{n,m}=e^{j2\pi\frac{r_{n-1}}{\lambda}\cos\frac{m-1}{M}\pi}$;
  \item $\mathbf{c}$ is the discrete reflectivity vector in the grids;
  \item $\gamma$ is a factor related to the noise level.
\end{itemize}

This $\ell _1$ norm minimization problem (\ref{cs}) can be solved by the iterative shrinkage thresholding algorithm (ISTA), a classic grid-based CS algorithm. Other mature CS recovery algorithms such as orthogonal matching pursuit (OMP) and compressive sampling matching pursuit (CoSaMP) can also be applied to this inversion problem.

\section{Proposed Method}\label{sec:method}
In this section, we introduce the concept of atomic norm and present an ANM-based DoA estimation algorithm for ULAs. Then we will extend it to NLAs by introducing the array manifold separation technique and the accelerated proximal gradient technique. Finally, a fast gridless DoA estimation algorithm based on ANM for NLAs is proposed.

\subsection{DoA via ANM for ULAs}

The grid-based CS algorithms assume that targets are located on discrete grids, which is an assumption that is inconsistent with the fact that the targets are usually off the grids, i.e., $\mathbf{a}\text{(}\theta_k\text{)}\notin \mathcal{A}_M$, resulting in the grid mismatch issue. ANM addresses this issue by manipulating $M\rightarrow \infty$, leveraging $\mathcal{A}=\left\{ \mathbf{a}\left( \theta \right) :\,\,\theta\in ( 0,\pi ) \right\} $.
Then, we can define the gridless version of Eq. \eqref{csl0} and Eq. \eqref{csl1} as
\begin{equation}
  \lVert \mathbf{x} \rVert _{\mathcal{A},0}=\text{inf}\left\{ K:\,\,\sum_{k=1}^K{\hat{c}_k\mathbf{a}\left( \hat{\theta}_k \right) =\mathbf{x,\,\,a}\left( \hat{\theta}_k \right) \in \mathcal{A}} \right\},
\end{equation}
\begin{equation}\label{anm}
  \begin{split}
    &\lVert \mathbf{x} \rVert _{\mathcal{A}} =\lVert \mathbf{x} \rVert_{\mathcal{A},1}                                                                                                                                              \\
    & =\text{inf}\left\{ \sum_{k=1}^K{\left| \hat{c}_k \right|}:\,\,\sum_{k=1}^K{\hat{c}_k\mathbf{a}\left( \hat{\theta}_k \right) =\mathbf{x,\,\,a}\left( \hat{\theta}_k \right) \in \mathcal{A}} \right\} .
  \end{split}
\end{equation}

Theoretically, $\left\{ \theta_k \right\}$ can be obtained by solving following optimization problem:
\begin{equation}\label{minanm}
  \min \lVert \mathbf{x} \rVert _{\mathcal{A}}.
\end{equation}

However, \eqref{minanm} is an infinite programming, which cannot be easily solved via mature convex optimization methods. \cite{TANGANM} has proved that when the atoms are in the form of Vandermonde vectors (i.e. it is a ULA with evenly distributed array elements), it can be solved via the following semi-positive definite programming:
\begin{equation}\label{sdp}
  \begin{array}{ll}
    \min_{\hat{\mathbf{x}},v,\mathbf{T}\left( \mathbf{u} \right)} & \tau(v+\frac{1}{N}\text{trace}\left( \mathbf{T}\left( \mathbf{u} \right) \right))+ \lVert \hat{\mathbf{x}}- \mathbf{x}\rVert_2^2 \\
    \text{s.t}.                                                   & \left( \begin{matrix}
                                                                               v                & \hat{\mathbf{x}}^{H}                \\
                                                                               \hat{\mathbf{x}} & \mathbf{T}\left( \mathbf{u} \right) \\
                                                                             \end{matrix} \right) \succeq 0;
  \end{array}
\end{equation}
where \begin{itemize}
  \item $\mathbf{T}( \mathbf{u} )$ is a Toeplitz matrix determined by its first row $\mathbf{u}$;
  \item $\tau$ is a factor related to the noise level.
\end{itemize}

Eq. \eqref{sdp} is a standard SDP that can be solved via popular convex optimization solvers. Angles $\left\{ \theta_k \right\}$ are embedded in $\mathbf{T}( \mathbf{u} )$ and can be revealed via Vandermonde decomposition, and $\left\{ c_k \right\}$ can be determined subsequentially without difficulty. The algorithm flow for gridless DoA estimation on ULAs by ANM is given in \algref{algulaanm}.

\begin{algorithm}[htbp]
  \SetAlgoLined
  \caption{DoA via ANM for ULAs}
  \label{algulaanm}

  \KwIn{$\mathbf{x}$, $\tau$, $K$.}
  \KwOut{$\{(\hat{c}_k,\hat{\theta}_k)\}$.}
  \BlankLine

  1. Solve the SDP problem Eq. \eqref{sdp};

  2. Performed eigenvalue decomposition on $\mathbf{T}\left( \mathbf{u} \right)$;

  3. Determine the noise subspace $\mathbf{U}_N$ based on the number of targets $K$;

  4. Define $\mathbf{p}(z)=
    \begin{bmatrix}
      1 & z & \cdots & z^{N-1}
    \end{bmatrix}^T$, $f(z) = \mathbf{p}^H(z)\mathbf{U}_N\mathbf{U}^H_N\mathbf{p}(z)$ and find the roots of the polynomial $f(z)$;

  5. Find the $K$ roots $\{z_k\}$ closest to the unit circle;

  6. Calculate $\{\hat{\theta}_k\}$ based on
  \begin{displaymath}
    \cos\hat{\theta}_k = \frac{\lambda}{2\pi d}\text{angle}(z_k),
    \nonumber
  \end{displaymath}
  where $\text{angle}(z_k)$ returns the phase angle of $z_k$ in the interval $[-\pi,\pi]$;

  7. The reflectivity $\{\hat{c}_k\}$ can be calculated by the least squares method.

\end{algorithm}

\subsection{DoA via Fast ANM for NLAs (FNLANM)}

ULAs with Vandermonde array manifold are the most well-studied array configuration. However, in practice, physical constraints such as limited space and mounting restrictions often make it challenging to employ ULAs. As mentioned earlier, NLAs offer improved angular resolution and reduced sidelobe levels with fewer elements. NLAs have provided a more practical and effective solution for many practical applications due to their superior performance. To take full advantage of gridless estimation, it is essential to extend the ANM-based DoA estimation approaches to NLAs.

\subsubsection{Manifold separation}

Inspired by the array interpolation technique, \cite{Jacobe} proposed the concept of the manifold separation technique. Using this method, the array manifold of a NLA can be decomposed into the product of a sampling matrix and a canonical Vandermonde manifold matrix. The sampling matrix describes the array itself and the Vandermonde matrix can be treated as the array manifold of a ULA, which can be used for DoA estimation.

The signal received by the $n$-th array element can be rewritten as $[\mathbf{a}(\theta)]_n$ as:
\begin{equation}\label{athetan}
  [\mathbf{a}(\theta)]_n=e^{\frac{1}{2}(2\pi \frac{r_{n-1}}{\lambda})(je^{j\theta}-\frac{1}{je^{j\theta}})}.
\end{equation}

By definition, the coefficients of the Laurent expansion of $e^{\frac{1}{2}z(t-\frac{1}{t})}$ are the Bessel functions of integer order $\{J_i(z)\}$:
\begin{equation}\label{bessel}
  e^{\frac{1}{2}z(t-\frac{1}{t})}=\Sigma_{i=-\infty}^{\infty} t^i J_i(z),
\end{equation}
where $J_{-i}(z)=(-1)^i J_i(z)$.

Performing a Laurent expansion of $[a(\theta)]_n$ according to Eq. \eqref{bessel}, we can express \eqref{athetan} as \cite{manifold4,manifold1,Jacobe}:
\begin{equation}
  [\mathbf{a}(\theta)]_n=\Sigma_{i=-\infty}^{\infty}j^iJ_i(2\pi \frac{r_{n-1}}{\lambda})e^{j\theta i}.
\end{equation}

One characteristic of the Bessel function is that it decays quickly with $|i|$. When $|i| >\frac{2\pi}{\lambda}r_{n}$, it can be shown that $J_i(2\pi \frac{r_n}{\lambda})$ is very small and  we can simply truncate the infinite series by ignoring the higher order terms. Note that the truncated order $I$ should satisfy the following condition:
\begin{equation}
  I>\frac{2\pi}{\lambda}D,
\end{equation}
where $D=\max \{r_n\}$\cite{Jacobe}.

Then the array manifold can be written as:
\begin{equation}
  \label{g}
  [\mathbf{a}(\theta)]_n\approx\Sigma_{i=-I}^{I}j^iJ_i(2\pi \frac{r_{n-1}}{\lambda})e^{j\theta i}=\mathbf{g}_{n-1}^T\mathbf{v}(\theta_k),
\end{equation}
where $\mathbf{g}_{n-1}=\begin{bmatrix}j^{-I}J_{-I}(2\pi \frac{r_{n-1}}{\lambda})\\\vdots\\J_0(2\pi \frac{r_{n-1}}{\lambda})\\\vdots\\j^{I}J_{I}(2\pi \frac{r_{n-1}}{\lambda})\end{bmatrix}$,
$\mathbf{v}(\theta)=\begin{bmatrix}e^{j\theta(-I)}\\\vdots\\1\\\vdots\\e^{j\theta(I)}\end{bmatrix}$.

Organizing $\mathbf{a}(\theta)$ into matrix form according to Eq. \eqref{g}, we obtain the following result of the array manifold separation as \cite{overview}:
\begin{equation}
  \label{G}
  \mathbf{a}(\theta)=\begin{bmatrix}\mathbf{g}_0^T\\\mathbf{g}_1^T\\\vdots\\\mathbf{g}_{N-1}^T\end{bmatrix}\mathbf{v}(\theta)=\mathbf{G}\mathbf{v}(\theta),
\end{equation}
where $\mathbf{G}$ is the $N\times(2I+1)$ sampling matrix which can be calculated from the real location of array elements or measured by effective aperture distribution function (EADF). Eq. \eqref{G} turns the array manifold of a NLA into Vandermonde with length $2I+1$, which eventually can be treated as the array manifold of a virtual ULA.

For noisy scenario with $K$ targets located at $K$ different directions, the signal $\mathbf{x}$ received by the array can be written as:
\begin{equation}
  \begin{array}{ll}
    \mathbf{x} & =\sum_{k=1}^K c_k \mathbf{a}(\theta_k)+\mathbf{n}            \\
               & =\sum_{k=1}^K c_k \mathbf{G}\mathbf{v}(\theta_k)+\mathbf{n}  \\
               & =\mathbf{G}\sum_{k=1}^K c_k \mathbf{v}(\theta_k)+\mathbf{n}.
  \end{array}
\end{equation}

Denote that $\mathbf{d}=\sum_{k=1}^K c_k\mathbf{v}(\theta_k)$, then $\mathbf{d}$ is the signal received by the virtual ULA and is the weighted sum of a series of Vandermonde vectors. As already mentioned in the previous section, for vectors of the form as $\mathbf{d}$, we can transform the atomic norm minimization problem into the following SDP:
\begin{equation}
  \label{eqsdpnla}
  \begin{array}{ll}
    \min_{\mathbf{d},v,\mathbf{T}\left( \mathbf{u} \right)} & \tau(v+\frac{1}{N}\text{trace}\left( \mathbf{T}\left( \mathbf{u} \right) \right))+ \lVert \mathbf{G}\mathbf{d}- \mathbf{x}\rVert_2^2 \\
    \text{s.t}.                                             & \left( \begin{matrix}
                                                                         v          & \mathbf{d}^{H}                      \\
                                                                         \mathbf{d} & \mathbf{T}\left( \mathbf{u} \right) \\
                                                                       \end{matrix} \right) \succeq 0.
  \end{array}
\end{equation}

The algorithm flow for gridless DoA estimation on NLAs by ANM (denoted as NLANM) is organized in \algref{nlaanmsdp}.

\begin{algorithm}[htbp]
  \SetAlgoLined 
  \caption{NLANM}
  \label{nlaanmsdp}

  \KwIn{$\mathbf{x}$, $\tau$, $K$, $\mathbf{r}$.}
  \KwOut{$\{(\hat{c}_k,\hat{\theta}_k)\}$.}
  \BlankLine

  1. Calculate the sampling matrix $\mathbf{G}$ based on Eq. \eqref{g} and Eq. \eqref{G};

  2. Solve the SDP problem of Eq. \eqref{eqsdpnla};

  3. Performed eigenvalue decomposition on $\mathbf{T}\left( \mathbf{u} \right)$;

  4. Determine the noise subspace $\mathbf{U}_N$ based on the number of targets $K$;

  5. Define $\mathbf{p}(z)=
    \begin{bmatrix}
      z^{-I} & \cdots & 1 & \cdots & z^{I}
    \end{bmatrix}^T$, $f(z) = \mathbf{p}^H(z)\mathbf{U}_N\mathbf{U}^H_N\mathbf{p}(z)$ and find the roots of the polynomial $f(z)$;

  6. Find the $K$ roots $\{z_k\}$ closest to the unit circle;

  7. Calculate $\{\hat{\theta}_k\}$ based on
  \begin{displaymath}
    \hat{\theta}_k = \text{abs}(\text{angle}(z_k)),
    \nonumber
  \end{displaymath}
  where $\text{abs}(\text{angle}(z_k))$ returns the absolute value of $\text{angle}(z_k)$;

  8. The reflectivity $\{\hat{c}_k\}$ can be calculated by the least squares method.

\end{algorithm}

\subsubsection{Acceleration method}

Based on the manifold separation technique, we successfully extended the ANM algorithm -- which can only be applied to ULAs -- to NLAs, and solved the problem by SDP. However, the feasibility of SDP is highly limited by the problem scale. In fact, the computational complexity of the SDP is $\mathcal{O}\left( N^{3.5} \right)$, where $N$ is the number of array elements \cite{sdp}. This is a huge computational load. When $N$ is larger than 32, it is usually inapplicable in practice. However, for the extension of the ANM algorithm to NLAs, we construct a new virtual array from the sampling matrix. The virtual array consists of $N_v=2I+1$ array elements, where $I>\frac{2\pi}{\lambda}D$. Thus, the complexity of SDP in \algref{nlaanmsdp} should be $\mathcal{O}(N_v^{3.5})$, which is usually unacceptable.

In this paper, inspired by \cite{ivdst}, we solve the above SDP problem \eqref{eqsdpnla} by APG and iterative shrinkage thresholding to reduce the solution complexity to $\mathcal{O}(N_v^{2})$.
APG can handle a variety of optimization problems with different constraints, including convex optimization problems with sparsity and low-rank constraints, by iteratively performing the three steps of smoothing, gradient descent, and approximate mapping. Considering the Vandermonde structure of $\mathbf{v}(\theta)$ and the sparsity of $\theta$, we add a threshold shrinkage step after gradient descent to guarantee the low-rank property of the Toeplitz matrix $\mathbf{T}\left( \mathbf{u} \right)$. The flow is shown in \algref{nlaanmapg}.

\begin{algorithm}[htbp]
  \SetAlgoLined 
  \caption{APG}
  \label{nlaanmapg}

  \KwIn{$\mathbf{x}$, $\mathbf{G}$, $\epsilon$, maxiter.}
  \KwOut{$\mathbf{T}\left( \mathbf{u} \right)$.}
  \BlankLine

  \textbf{Initialize:} \begin{itemize}
    \item $i=1$, $t_0=1$, $\mathbf{d}^0=\mathbf{G}^H\mathbf{x}$;
    \item $\mathbf{T}_0=\mathcal{T}(\mathbf{d}_0\mathbf{d}_0^H)$, where $\mathcal{T}(\mathbf{d}_0\mathbf{d}_0^H)$ returns a Toeplitz matrix by averaging all
          elements along each diagonal direction of $\mathbf{d}_0\mathbf{d}_0^H$;
    \item $v_0=\text{trace}(\mathbf{T}_0)/N_v$,
    \item $\boldsymbol{f}_1=\boldsymbol{f}_0=(\mathbf{d}_0,v_0,\mathbf{T}_0)$.
  \end{itemize}

  \textbf{repeat}

  1. Set $\bar{\boldsymbol{f}_i}=\boldsymbol{f}_i+\frac{t_{i-1}-1}{t_i}(\boldsymbol{f}_i-\boldsymbol{f}_{i-1})=(\bar{\mathbf{d}_i},\bar{v_i},\bar{\mathbf{T}_i})$, where $t_i=\frac{1+\sqrt{4t_{i-1}^2+1}}{2}$;

  2. Calculate $\mathbf{d}_g=\bar{\mathbf{d}_i}-\delta \mathbf{G}^H(\mathbf{G}\bar{\mathbf{d}_i}-\mathbf{x})$, $v_g=\bar{v_i}$ and $\mathbf{T}_g=\bar{\mathbf{T}_i}$, where $\delta$ is an appropriate small stepsize;

  3. Decompose $\mathbf{T}_g$ into $\mathbf{T}_g=\mathbf{V}\boldsymbol{\Lambda}\mathbf{V}^H$;

  4. Shrink $\boldsymbol{\Lambda}$ to $\tilde{\boldsymbol{\Lambda}}$ and set $\tilde{\mathbf{T}_g}=\mathbf{V}\tilde{\boldsymbol{\Lambda}}\mathbf{V}^H$;

  5. Set $\mathbf{Z}=[\text{trace}(\tilde{\boldsymbol{\Lambda}})\mathbf{d}_g^H; \mathbf{d}_g\tilde{\mathbf{T}_g}]$ and decompose $(\boldsymbol{\Sigma},\mathbf{U})=\text{eigs}(\mathbf{Z},\text{rank}(\tilde{\boldsymbol{\Lambda}})+1)$;

  6. Set $\tilde{\mathbf{Z}}=\mathbf{U}\boldsymbol{\Sigma}\mathbf{U}^H$;

  7. Update $\boldsymbol{\theta}_{i+1}=(\mathbf{d}_{i+1},v_{i+1},\mathbf{T}_{i+1})$, where $\mathbf{d}_{i+1}=\tilde{\mathbf{Z}}_{2:end,1}$, $v_{i+1}=\tilde{\mathbf{Z}}_{1,1}$, and $\mathbf{T}_{i+1}=\tilde{\mathbf{Z}}_{2:\text{end},2:\text{end}}$;

  8. $i=i+1$;

  \textbf{until} $i>$ maxiter or $\lVert \mathbf{T}_{i+1}-\mathbf{T}_{i} \rVert_F\leq\epsilon$

  $\mathbf{T}\left( \mathbf{u} \right)=\mathbf{T}_{i+1}$.

\end{algorithm}

The proposed method is named as the FNLANM algorithm (from the acrostic fast NLANM), which is a fast implementation of the NLANM algorithm. This proposed method achieves accelerated gridless DoA estimation for arbitrary linear arrays. The detailed algorithm flow is given in \algref{nlaanmivdst}, and the computational complexity is given in \tabref{com}.

\begin{algorithm}[htbp]
  \SetAlgoLined 
  \caption{FNLANM}
  \label{nlaanmivdst}

  \KwIn{$\mathbf{x}$, $\tau$, $K$, $\mathbf{r}$, $\epsilon$, maxiter.}
  \KwOut{$\{(\hat{c}_k,\hat{\theta}_k)\}$.}
  \BlankLine

  1. Calculate the sampling matrix $\mathbf{G}$ based on Eq. \eqref{G};

  2. $\mathbf{T}\left( \mathbf{u} \right)$ = APG($\mathbf{x}$, $\mathbf{G}$, $\epsilon$, maxiter);

  3. Performed eigenvalue decomposition on $\mathbf{T}\left( \mathbf{u} \right)$;

  4. Determine the noise subspace $\mathbf{U}_N$ based on the number of targets $K$;

  5. Define $\mathbf{p}(z)=
    \begin{bmatrix}
      z^{-I} & \cdots & 1 & \cdots & z^{I}
    \end{bmatrix}^T$, $f(z) = \mathbf{p}^H(z)\mathbf{U}_N\mathbf{U}^H_N\mathbf{p}(z)$ and find the roots of the polynomial $f(z)$;

  6. Find the $K$ roots $\{z_k\}$ closest to the unit circle;

  7. Calculate $\{\hat{\theta}_k\}$ based on
  \begin{displaymath}
    \hat{\theta}_k = \text{abs}(\text{angle}(z_k)),
    \nonumber
  \end{displaymath}
  where $\text{abs}(\text{angle}(z_k))$ returns the absolute value of $\text{angle}(z_k)$;

  8. The reflectivity $\{\hat{c}_k\}$ can be calculated by the least squares method.

\end{algorithm}

\begin{table}[htbp]
  \begin{center}
    \caption{Computational complexity.}
    \label{com}
    \begin{tabular}{ c  c c c}
      \toprule
      Algorithm                & ANM                                 & NLANM & FNLANM \\
      \midrule
      Computational complexity & $\mathcal{O}\left( N^{3.5} \right)$
                               & $\mathcal{O}(N_v^{3.5})$
                               & $\mathcal{O}(N_v^{2})$                               \\
      \bottomrule
    \end{tabular}
  \end{center}
\end{table}

\section{Simulation Experiments}\label{sec:simu}
\subsection{Performance metrics}
\subsubsection{Accuracy}
We measure the accuracy of DoA estimation in terms of the root mean square error (RMSE) between the DoA estimated result $\hat{\boldsymbol{\theta}}$ and the ground truth $\boldsymbol{\theta_0}$. RMSE is defined as follows
\begin{equation}
  \text{RMSE} = \sqrt{\frac{\sum_{i=1}^{MC}\|\hat{\boldsymbol{\theta}}_i-\boldsymbol{\theta}_0\|_2^2}{MC\times K}},
\end{equation}
where $MC$ is the number of Monte Carlo experiments and $K$ is the number of targets.
\subsubsection{Robustness}
We measure the robustness of DoA estimation in terms of the success rate, which is defined as follows
\begin{equation}\label{sr}
  \text{SR}= \frac{mc_t}{MC},
\end{equation}
where $mc_t$ is the total amount of experiments that satisfy $\max|\hat{\boldsymbol{\theta}}-\boldsymbol{\theta}_0|<\gamma$ and $\gamma$ is a constant small enough.

\subsection{Non-uniform linear array}
In this subsection, we aim to prove the necessity of extending ANM to NLAs by setting arrays with different location deviations (LD). We compare the NLA and ULA with the same aperture $D$ and the number of elements $N$, and define the location deviation as follows:
\begin{equation}
  \text{LD} = \sqrt{\frac{\|\mathbf{r}-\mathbf{r}_{ula}\|_2^2}{N}}/d,
\end{equation}
where \begin{itemize}
  \item $\mathbf{r}=
          \begin{bmatrix}
            0 & r_1 & \cdots & r_{N-1}
          \end{bmatrix}^T$ is the position of each array element of the NLA;
  \item $\mathbf{r}_{ula}=
          \begin{bmatrix}
            0 & d & \cdots & (N-1)d
          \end{bmatrix}^T$ is the position of each array element of the ULA.
\end{itemize}

The influence of location deviation on the estimation accuracy of the three ANM-based algorithms is discussed through multiple Monte Carlo experiments. In this experiment, we increase location deviation progressively and other simulation parameters are shown in \tabref{tabLD}.

\begin{table}[htbp]
  \begin{center}
    \caption{Simulation parameters.}
    \label{tabLD}
    \begin{tabular}{ c  c  c }
      \toprule
      Parameters               & Symbols & Values   \\
      \midrule
      Frequency                & $f$     & 77.5 GHz \\
      Number of array elements & $N$     & 16       \\
      Aperture                 & $D$     & 29.0 mm  \\
      Signal-to-noise ratio    & SNR     & 30 dB    \\
      \bottomrule
    \end{tabular}
  \end{center}
\end{table}

\begin{figure}[htbp]
  \centering
  \includegraphics[width=0.6\linewidth]{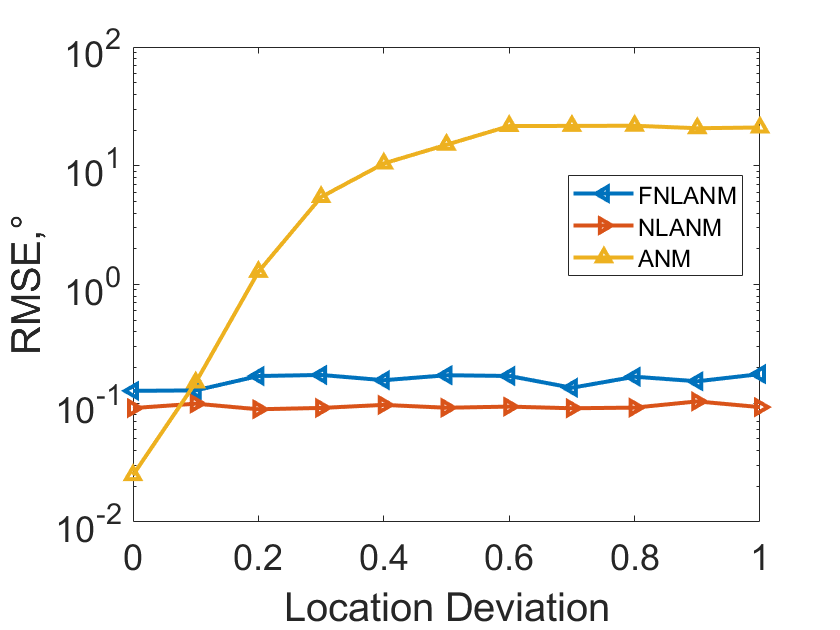}
  \caption{Estimation accuracy comparison of ANM, NLANM and FNLANM versus location deviation.}
  \label{nlarmseld}
\end{figure}

At each location deviation, 1000 different arrays are randomly generated and DoA estimation is performed by ANM, NLANM and FNLANM algorithms. The RMSEs of the three algorithms are shown in \figref{nlarmseld}. It can be seen that the RMSE of ANM increases rapidly with the increase of LD. When LD reaches 0.3, the RMSE of ANM approaches $10^{\circ}$, which means that in almost all tested arrays of $LD=0.3$, the difference between the DoA estimated result and the ground truth is large. Such extremely unreliable estimates cannot be applied in practice. However, the RMSEs of NLANM and FNLANM are not affected by the location deviation.
The complexities and the average computation times of DoA estimation algorithms are given in \tabref{tabLDtime}. For computation, Intel CPU i5-8265U (1.60 GHz) and 8 GB RAM are used. It is worth mentioning that FNLANM greatly reduces the average runtime while maintaining almost the same DoA estimation accuracy as NLANM.

\begin{table}[htbp]
  \begin{center}
    \caption{Average runtime.}
    \label{tabLDtime}
    \begin{tabular}{ c  c  c c}
      \toprule
      Algorithm         & ANM & NLANM & FNLANM \\
      \midrule
      Average runtime,s & 1.7 & 34.3  & 6.0    \\
      \bottomrule
    \end{tabular}
  \end{center}
\end{table}

This statistical experiment demonstrates the necessity of extending the existing frame of the ANM algorithm to NLAs and accelerating the algorithm.

\subsection{Off-grid effect}
All the following simulation experiments are based on NLAs. The simulation parameters are shown in the \tabref{taboff}, where the super-resolution factor is a constant relative to the grid interval of grid-based CS algorithms. The theoretical resolution of the system $\rho$ and the grid interval $\rho_s$ can be calculated according to Eq. (\ref{rho}) and Eq. (\ref{rhos}) \cite{review2}.

\begin{table}[htbp]
  \begin{center}
    \caption{Simulation parameters.}
    \label{taboff}
    \begin{tabular}{ c  c  c }
      \toprule
      Parameters               & Symbols & Values   \\
      \midrule
      Frequency                & $f$     & 77.5 GHz \\
      Number of array elements & $N$     & 16       \\
      Aperture                 & $D$     & 29.0 mm  \\
      Location deviation       & $LD$    & 0.3      \\
      Signal-to-noise ratio    & SNR     & 20 dB    \\
      Super-resolution factor  & $\eta$  & 4        \\
      \bottomrule
    \end{tabular}
  \end{center}
\end{table}

\begin{equation}
  \label{rho}
  \rho = 1.22\frac{\lambda}{D}
\end{equation}

\begin{equation}
  \label{rhos}
  \rho_s = \frac{\rho}{\eta}
\end{equation}

\begin{figure}[htbp]
  \centering
  \subfloat[]{\label{0single}\includegraphics[width=0.45\linewidth]{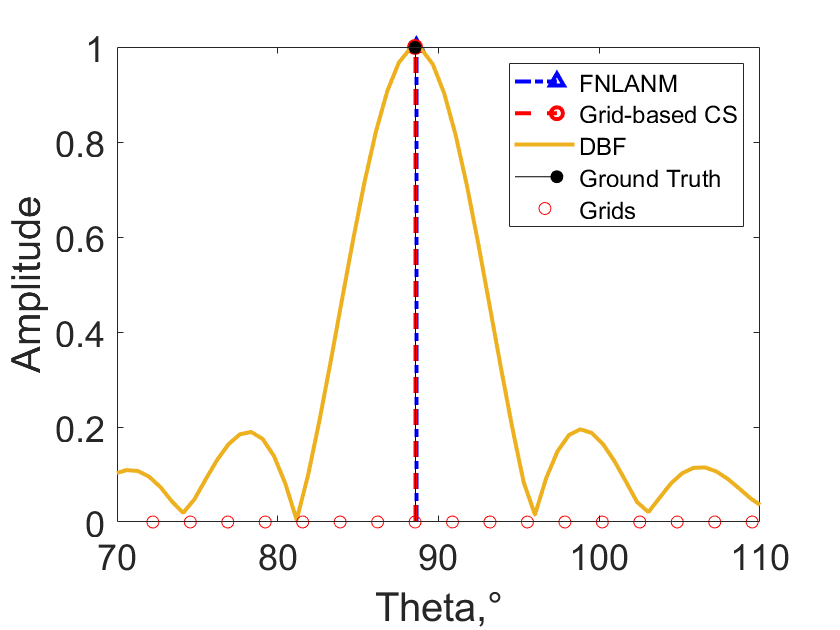}}
  \hfill
  \subfloat[]{\label{16single}\includegraphics[width=0.45\linewidth]{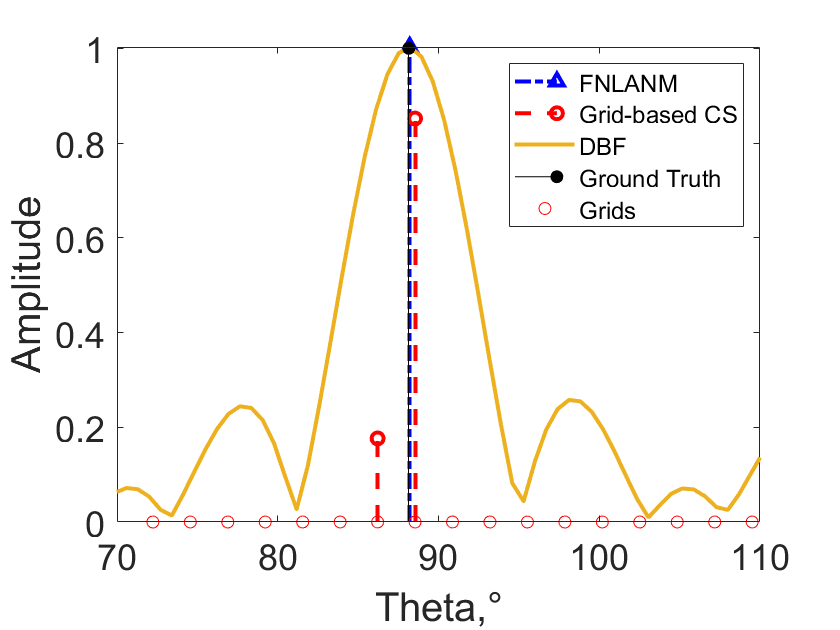}}
  \vfill
  \subfloat[]{\label{26single}\includegraphics[width=0.45\linewidth]{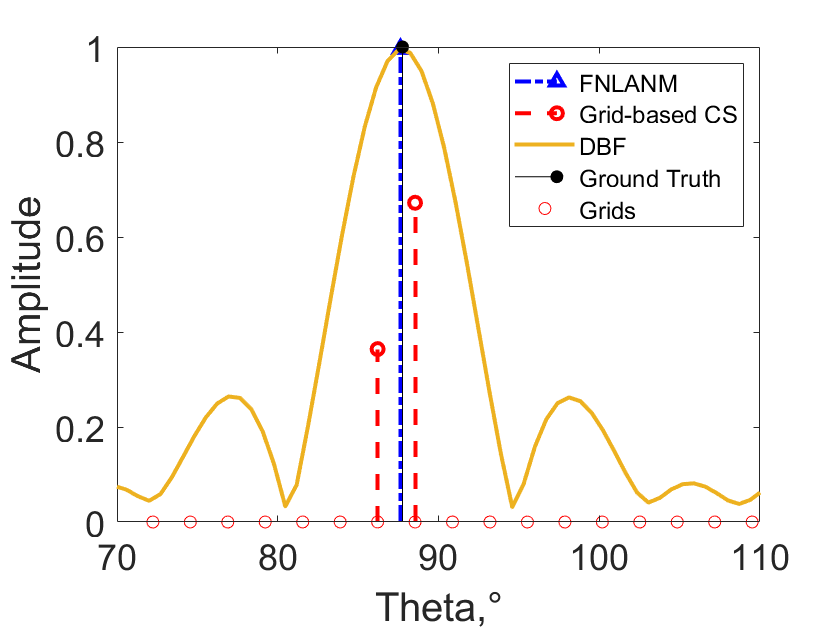}}
  \hfill
  \subfloat[]{\label{24single}\includegraphics[width=0.45\linewidth]{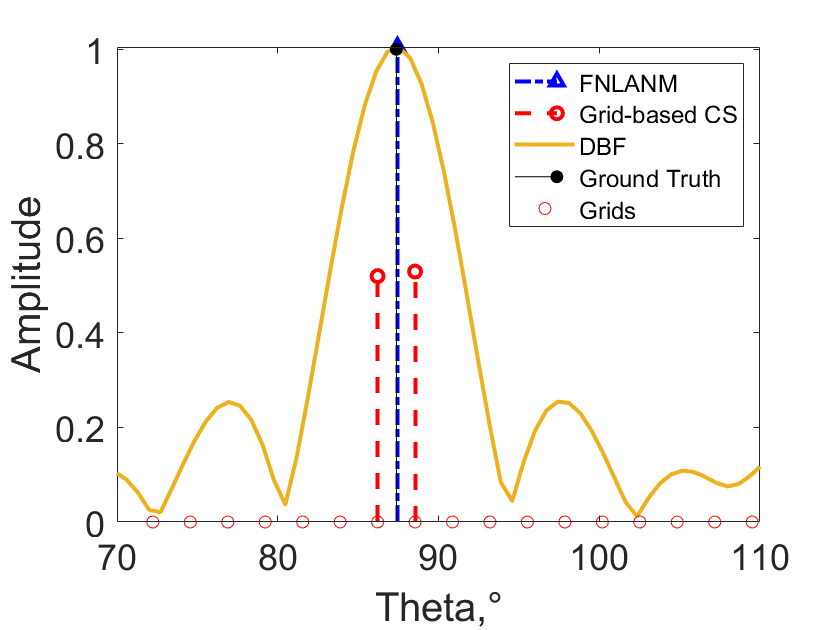}}
  \vfill
  \caption{Comparison of estimation results of DBF, grid-based CS and FNLANM. The off-grid error of the single target is (a) 0, (b) $\rho_s/6$, (c) $\rho_s/3$, (d) $\rho_s/2$.}
  \label{offgriderror}
\end{figure}

The DoA estimation results of DBF, grid-based CS and the proposed algorithm FNLANM are compared by setting a single target at different positions. By comparing the results in \figref{offgriderror}, we can see that: for the grid-based CS algorithm, when the target is off the grid, the energy of the real target will leak to the nearby grid, resulting in the difference between the predicted sparsity and the real sparsity, causing false alarms and missed alarms. This effect becomes more severe as the off-grid error becomes larger. However, DBF and FNLANM are completely unaffected by the off-grid effect. With this simple example, we show the impact of the off-grid effect on the DoA estimation performance of grid-based CS algorithms, which in turn illustrates the necessity of research on gridless DoA estimation methods.

\begin{figure}[htbp]
  \centering
  \subfloat[]{\includegraphics[width=0.6\linewidth]{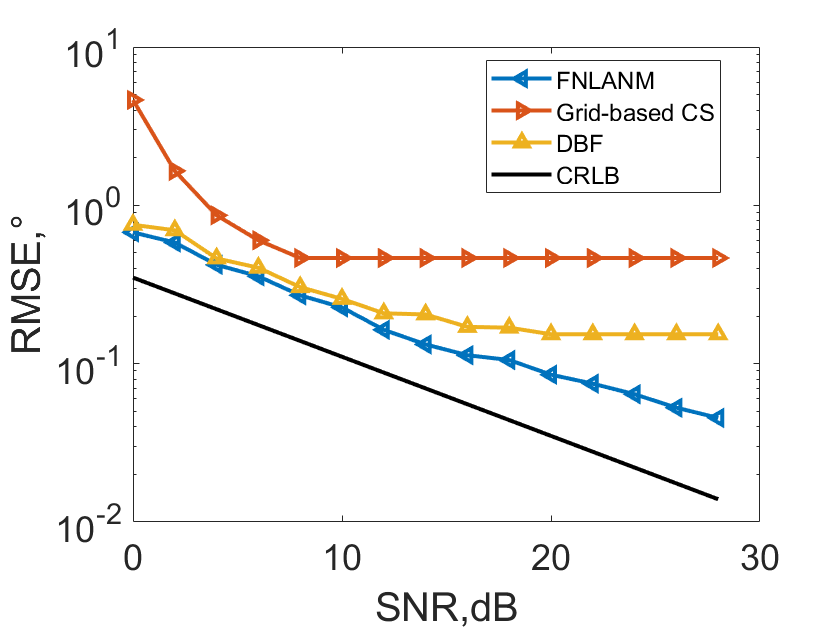}}
  \vfill
  \subfloat[]{\includegraphics[width=0.6\linewidth]{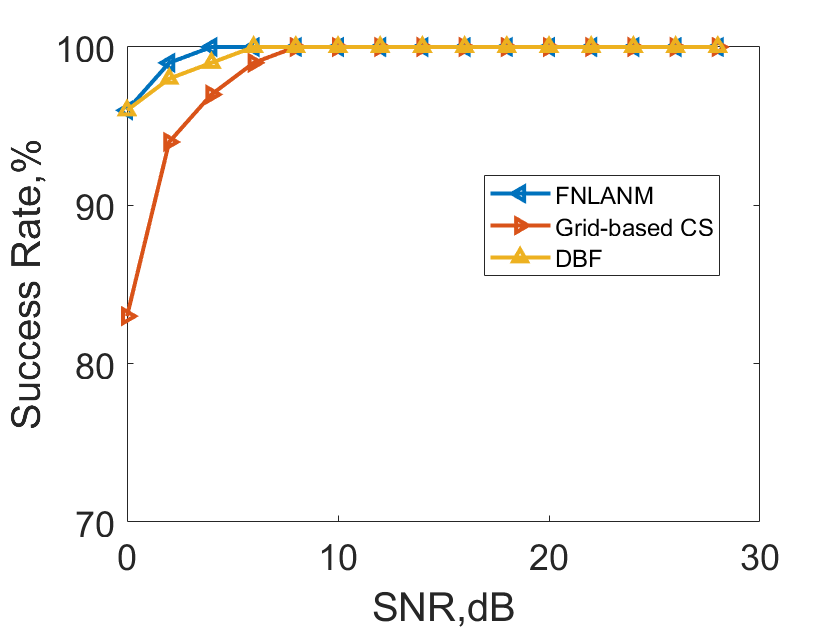}}
  \caption{Estimation performance of DBF, grid-based CS and FNLANM for an off-grid target versus the SNR. (a) RMSE. (b) SR for $\gamma=\rho_s/2$.}
  \label{pdsingle}
\end{figure}

Next, we investigate the estimation performance of three algorithms for an off-grid target through Monte Carlo experiments. The RMSEs of the three algorithms are compared with the Cram\'er-Rao lower bound (CRLB). We set $\gamma=\rho_s/2$ and calculate the success rates of the three algorithms according to Eq. \eqref{sr}. As seen in \figref{pdsingle}, the estimation performance of all three algorithms becomes better with increasing SNR with the same trend. As the SNR rises to 8 dB, the RMSE of grid-based CS decreases rapidly and then plateaus, eventually stabilizing at a constant associated with the off-grid error of the target. However, the RMSEs of ANM and DBF are almost the same when the SNR is less than 10 dB, and are much better than the grid-based CS algorithm for all tested SNRs. When the SNR is large than 20 dB, the RMSE decline of the DBF algorithm becomes flat and almost no longer decreases with increasing SNR, while the ANM is always approaching the CRLB.

From this experiment, we can conclude that for the DoA estimation of an off-grid target, the accuracy and robustness of the FNLANM algorithm outperform the grid-based CS algorithm.

\subsection{Number of array elements}

In this part of the experiment, we investigate the effect of the number of array elements $N$ on the DoA estimation performance for a certain array aperture $D$. To demonstrate more clearly, a variable is defined as
\begin{equation}
  \alpha = \frac{N}{N_{ula}},
\end{equation}
where $N_{ula}$ is the number of array elements of a standard ULA satisfying $D=\frac{(N_{ula}-1)\lambda}{2}$.

\begin{figure}[htbp]
  \centering
  \subfloat[]{\includegraphics[width=0.6\linewidth]{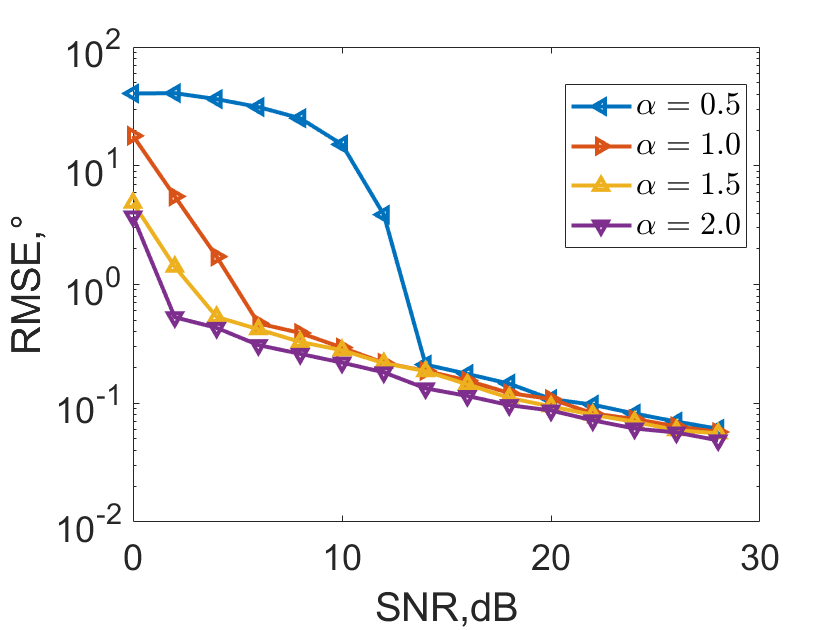}}
  \vfill
  \subfloat[]{\includegraphics[width=0.6\linewidth]{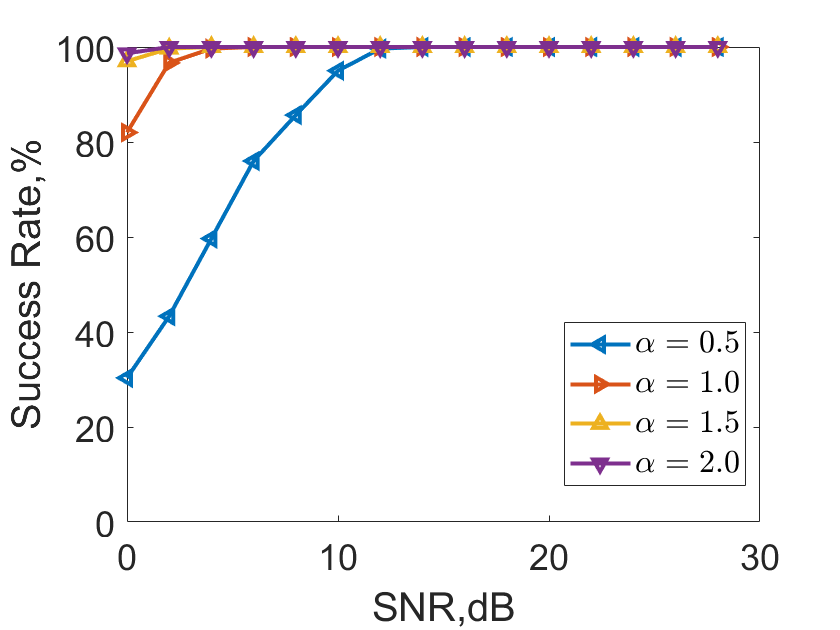}}
  \caption{Estimation performance of FNLANM for different $\alpha$ versus the SNR. (a) RMSE. (b) SR for $\gamma=\rho/4$.}
  \label{sensimuN}
\end{figure}

We simulate $N_{ula}= 16$, $D=29.0$ mm, $LD=0.3$, and $\alpha=\{0.5, 1.0, 1.5, 2.0\}$ respectively, and evaluate the estimation performance as a function of the SNR. The experimental results of DoA estimation by FNLANM for different $\alpha$ are shown in \figref{sensimuN}, and the average runtimes are shown in \tabref{tabsimuN}. As seen in the results, the estimation performance all becomes better with increasing SNR with the same trend for different $\alpha$. When $\alpha=0.5$, a SNR greater than 14 dB is required to achieve a valid and stable DoA estimate. When $\alpha$ increases from 0.5 to 1.0, both RMSE and SR gain significant improvements, especially at SNR below 14 dB. However, when $\alpha$ continues to increase, the enhancement of both evaluation metrics is no longer significant. It is interesting to mention that the number of array elements $N$ has almost no effect on the average runtime with a constant array aperture $D$.

\begin{table}[htbp]
  \begin{center}
    \caption{Average runtime.}
    \label{tabsimuN}
    \begin{tabular}{ c  c  c c  c}
      \toprule
      $\alpha$          & 0.5 & 1.0 & 1.5 & 2.0 \\
      \midrule
      Average runtime,s & 5.9 & 6.0 & 6.0 & 6.1 \\
      \bottomrule
    \end{tabular}
  \end{center}
\end{table}

This experiment illustrates that for a certain array aperture, it is not necessary for FNLANM to pursue a dense array element arrangement to achieve a better DoA estimation performance, which helps to reduce the amount of data collection.

\subsection{Array aperture}

In this part of the experiments, we keep $\alpha$ constant and investigate the effect of array aperture $D$ on the DoA estimation performance.

\begin{figure}[htbp]
  \centering
  \subfloat[]{\includegraphics[width=0.6\linewidth]{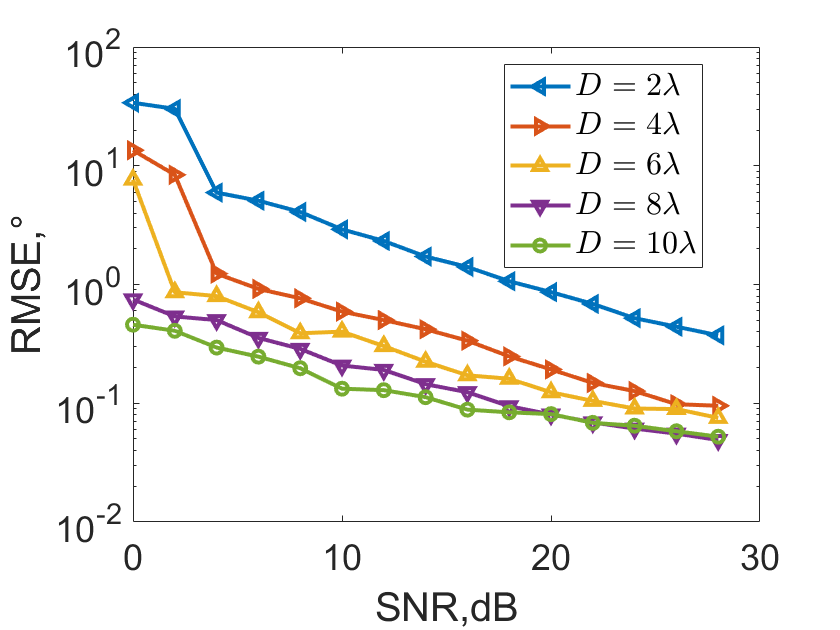}}
  \vfill
  \subfloat[]{\includegraphics[width=0.6\linewidth]{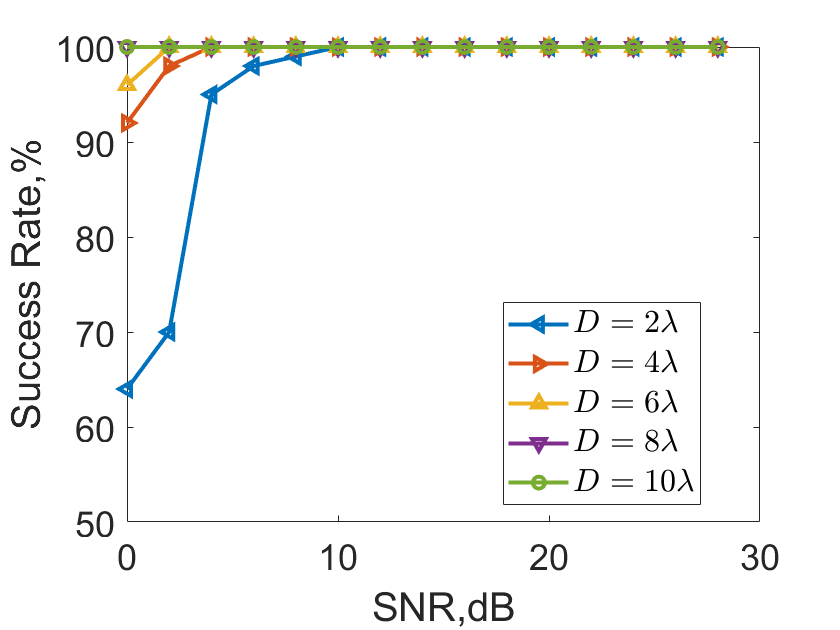}}
  \vfill
  \subfloat[]{\label{NDtime}\includegraphics[width=0.6\linewidth]{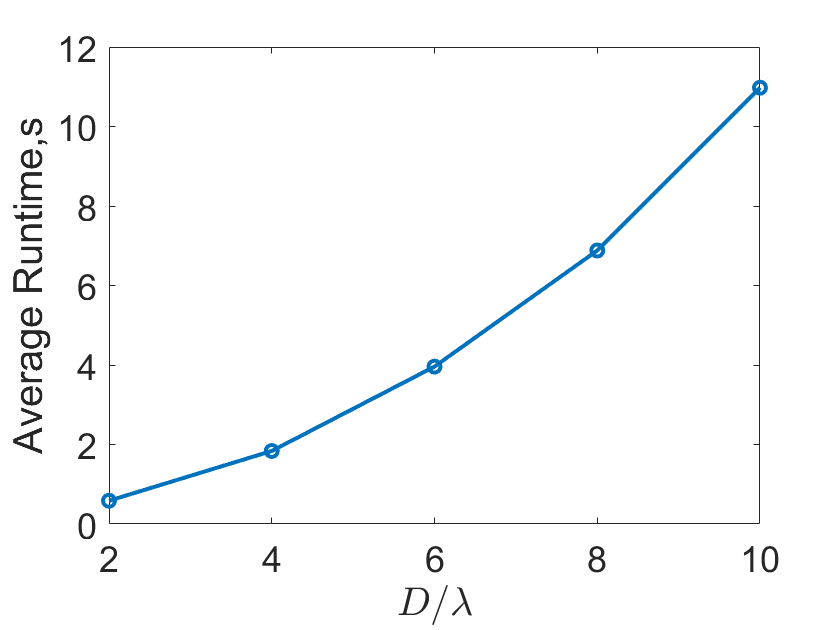}}
  \caption{Estimation performance of FNLANM for different $D$. (a) RMSE versus the SNR. (b) SR versus the SNR for $\gamma=\rho/4$ (c) Average runtime.}
  \label{sensimuD}
\end{figure}

We simulate $\alpha=1.0$, $LD=0.3$ and $D=\{2, 4, 6,8, 10\}\lambda$ respectively, and evaluate the estimation performance as a function of the SNR. The experimental results of DoA estimation by FNLANM for different $D$ are shown in \figref{sensimuD}. From the results, it can be seen that the different $D$ have the same tendency to become better in estimation performance as the SNR increases. When $D=2\lambda$, a SNR greater than 10 dB is required to obtain a valid and stable DoA estimate for a single target. When $D$ is increased from $2\lambda$ to $4\lambda$, a significant improvement of RMSE is obtained at all tested SNRs. However, as $D$ continues to increase, the RMSE still improves, but the improvement becomes progressively smaller. The effect of $D$ on SR is mainly reflected when the SNR is below 10 dB. A larger array aperture can achieve a higher success rate of DoA estimation at low SNR. However, it can be seen from \figref{NDtime} that although the proposed method FNLANM is already an improved algorithm after accelerating the NLANM algorithm by APG, its average runtime still rises rapidly when the array aperture $D$ increases.

This experiment illustrates that it is not necessary for FNLANM to pursue massive arrays to obtain better DoA estimation performance, which helps to reduce the hardware requirements.

\subsection{Angular super-resolution}

Theoretically, when the distance between two targets is larger than the theoretical resolution $\rho$, DBF, grid-based CS and ANM all can separate the two targets; when the distance between two targets is smaller than the theoretical resolution $\rho$, DBF cannot separate them, while both grid-based CS and ANM have super-resolution capability to separate the two targets.
We simulate $K=2$ targets with $\theta_1\in[0,\pi]$ and $\theta_2 = \theta_1-\Delta\theta$ in this experiment and
evaluate the estimation performance as a function of $\Delta\theta$.

\begin{figure}[htbp]
  \centering
  \subfloat[]{\label{nlaingriddoublerho}\includegraphics[width=0.45\linewidth]{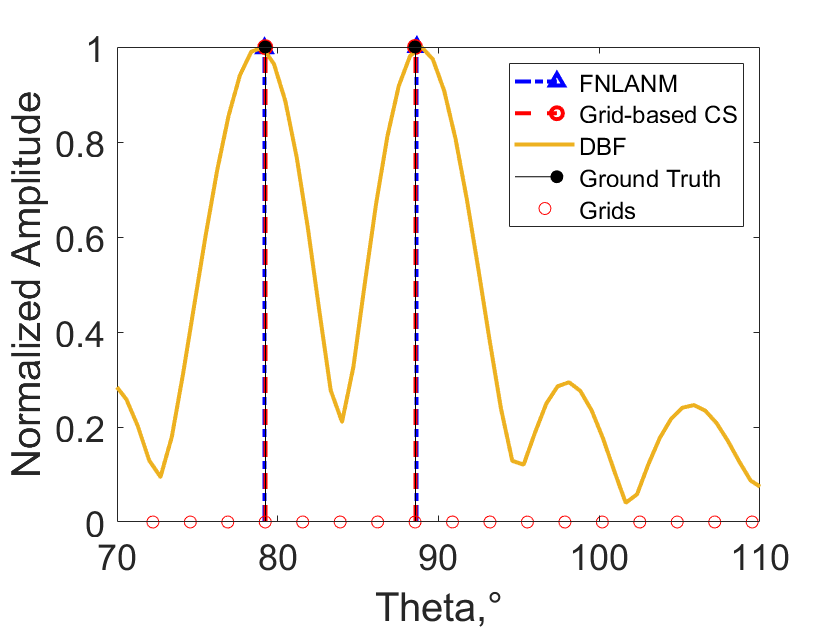}}\hfill
  \subfloat[]{\label{nlaoffgriddoublerho}\includegraphics[width=0.45\linewidth]
    {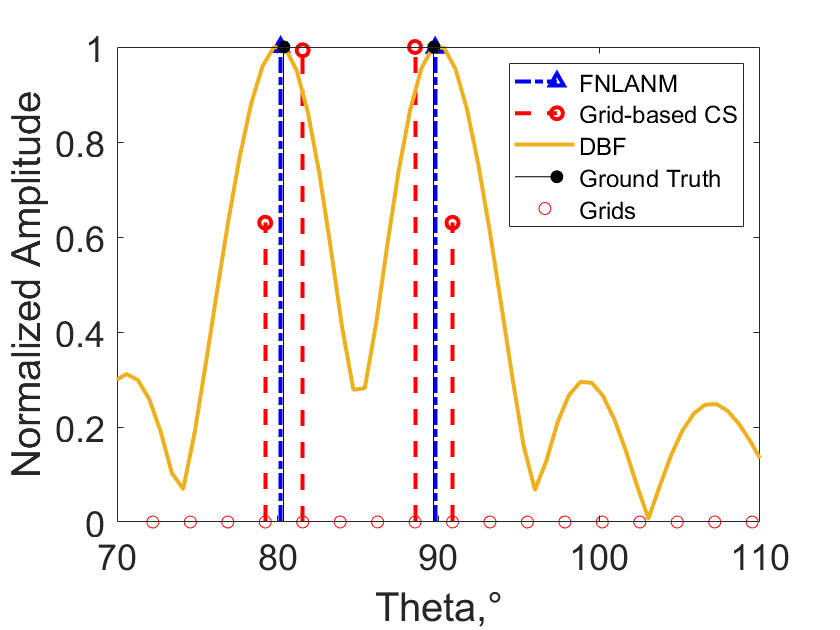}}\vfill
  \subfloat[]{\label{nlaingriddoublerhos}\includegraphics[width=0.45\linewidth]{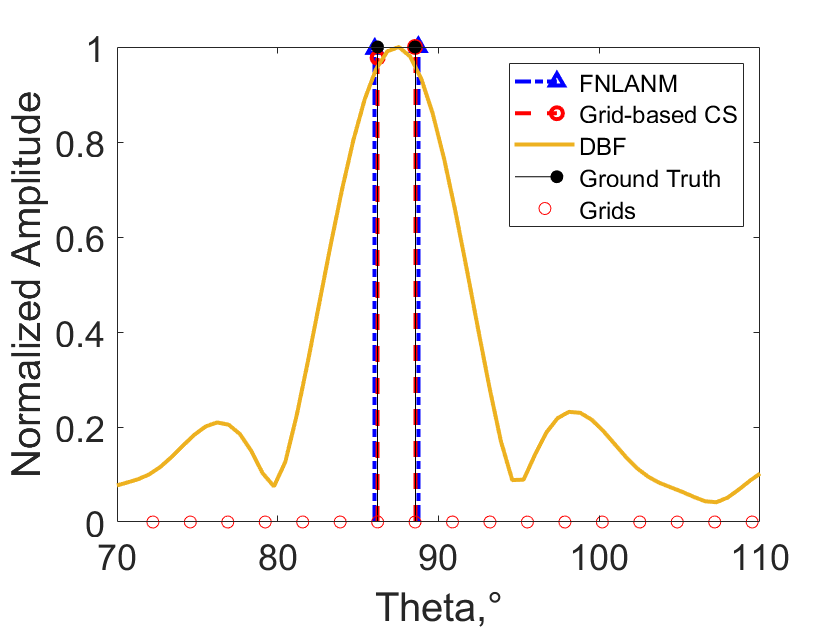}}\hfill
  \subfloat[]{\label{nlaoffgriddoublerhos}\includegraphics[width=0.45\linewidth]{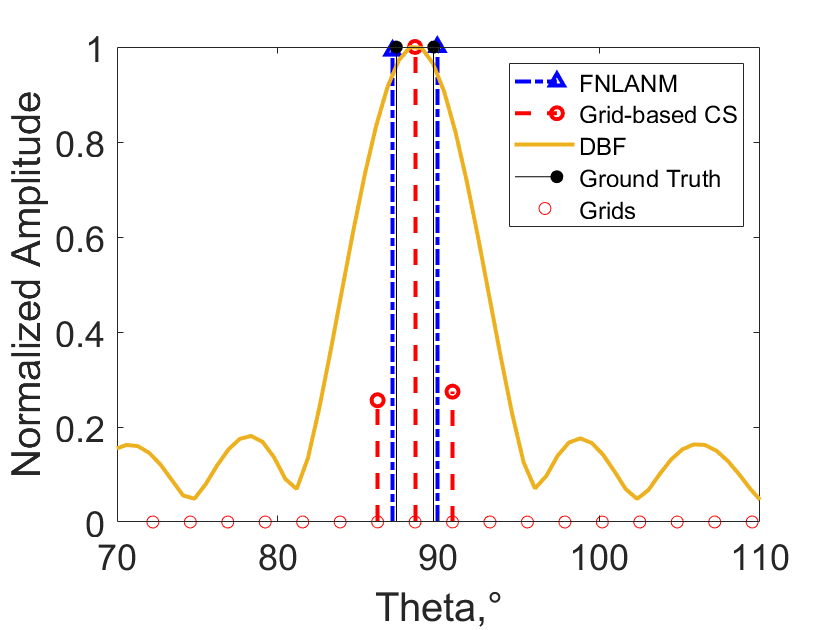}}
  \caption{Comparison of reconstruction results by DBF, grid-based CS and FNLANM. (a) On-grid targets with $\Delta\theta=\rho$. (b) Off-grid targets with $\Delta\theta=\rho$. (c) On-grid targets with $\Delta\theta=\rho_s$. (d) Off-grid targets with $\Delta\theta=\rho_s$.}
  \label{theta4}
\end{figure}

Setting $\Delta\theta = \{\rho,\rho_s\}$ and setting $\theta_1$ off-grid and on-grid respectively, the experimental results are shown in \figref{theta4}.
By comparing \figref{nlaingriddoublerho} and \figref{nlaoffgriddoublerho}, we can see that when the angular separation between two targets is large enough, whether the targets are off-grid or not only affects the reconstruction performance of the grid-based CS algorithm, while both DBF and FNLANM can accurately locate two targets.
By comparing \figref{nlaingriddoublerho} and \figref{nlaingriddoublerhos}, we can see that when the angular separation of two on-grid targets is less than the theoretical resolution $\rho$, both the grid-based CS algorithm and FNLANM exhibit super-resolution capability to separate the targets.
Comparing \figref{nlaingriddoublerhos} and \figref{nlaoffgriddoublerhos}, we can see that the super-resolution capability of grid-based CS is severely affected by the off-grid effect, and only FNLANM among the three algorithms can accurately locate two targets when the angular separation between two off-grid targets is less than the theoretical resolution.

\begin{figure}[htbp]
  \centering
  \includegraphics[width=0.6\linewidth]{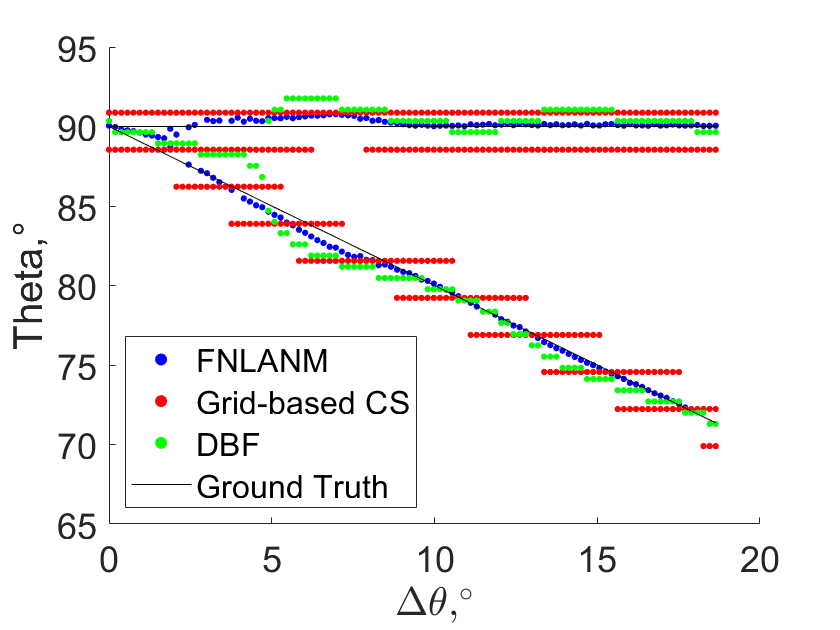}
  \caption{Comparison of estimation results by DBF, grid-based CS and FNLANM versus angular separation for $N=16$, $D=29.0$ mm, $\text{SNR}=20$ dB.}
  \label{deltatheta}
\end{figure}

With the simulation parameters as shown in \tabref{taboff}, setting $\theta_1=90^{\circ}$ and gradually increasing $\Delta\theta$ from 0 to $2\rho$, the DoA estimation results are shown in \figref{deltatheta}. The results show that with the theoretical resolution $\rho=9.32^{\circ}$, FNLANM can separate the targets that are $2.45^{\circ}$ apart, which demonstrates a good super-resolution capability. However, the super-resolution capability of the grid-based CS algorithm is severely deteriorated because of the off-grid effect.

\begin{figure}[htbp]
  \centering
  \subfloat[]{\includegraphics[width=0.6\linewidth]{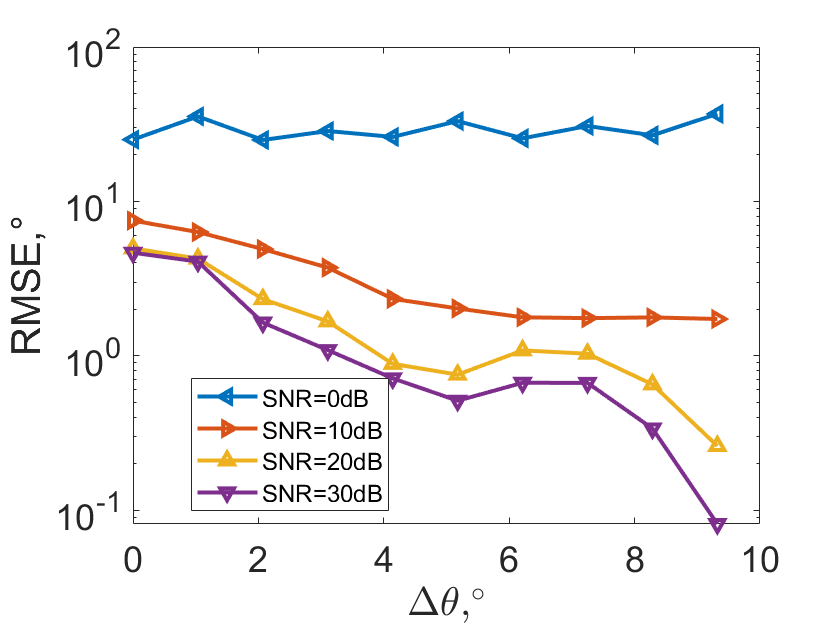}}
  \hfill
  \subfloat[]{\includegraphics[width=0.6\linewidth]{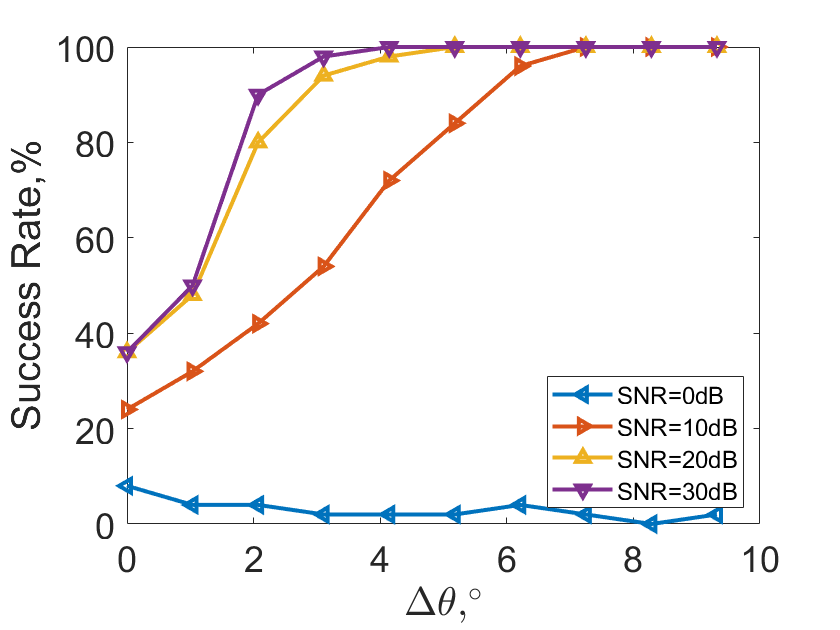}}
  \caption{Estimation performance of FNLANM for different SNRs versus angular separation. (a) RMSE. (b) SR for $\gamma=\rho/4$.}
  \label{thetaSNR}
\end{figure}

To better investigate the super-resolution ability of FNLANM, we examine the variation of RMSE and SR with $\Delta\theta\in(0,\rho)$ under different SNRs by statistical experiments, and the results are shown in \figref{thetaSNR}. The experimental results conform to the expectation that the estimation performance of FNLANM becomes better as the SNR increases. When the $\text{SNR} = 0$ dB, FNLANM can achieve almost unreliable DoA estimation. When the SNR rises to 10 dB, FNLANM can stably separate targets with an interval slightly smaller than $\rho$. As the SNR rises to 20 dB, FNLANM can separate targets spaced approximately $\rho/3$ apart. As the SNR continues to rise, the minimum $\Delta\theta$ that can be separated hardly changes anymore.

To summarize, FNLANM has super-resolution capability with a resolution of $\rho/3$ at a sufficiently high SNR. The experiments in this subsection illustrate the superiority of FNLANM in terms of super-resolution.

\subsection{Automotive scenario}

\begin{figure}[htbp]
  \centering
  \subfloat[]{\label{simuscene}\includegraphics[width=0.6\linewidth]{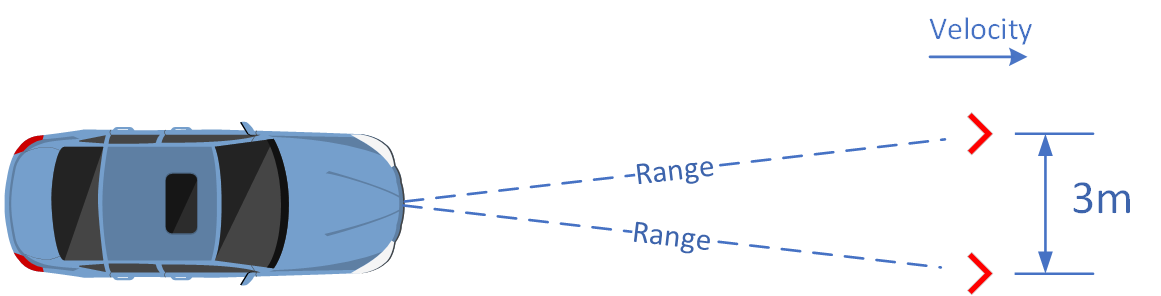}}
  \vfill
  \subfloat[]{\label{1050simu}\includegraphics[width=0.6\linewidth]{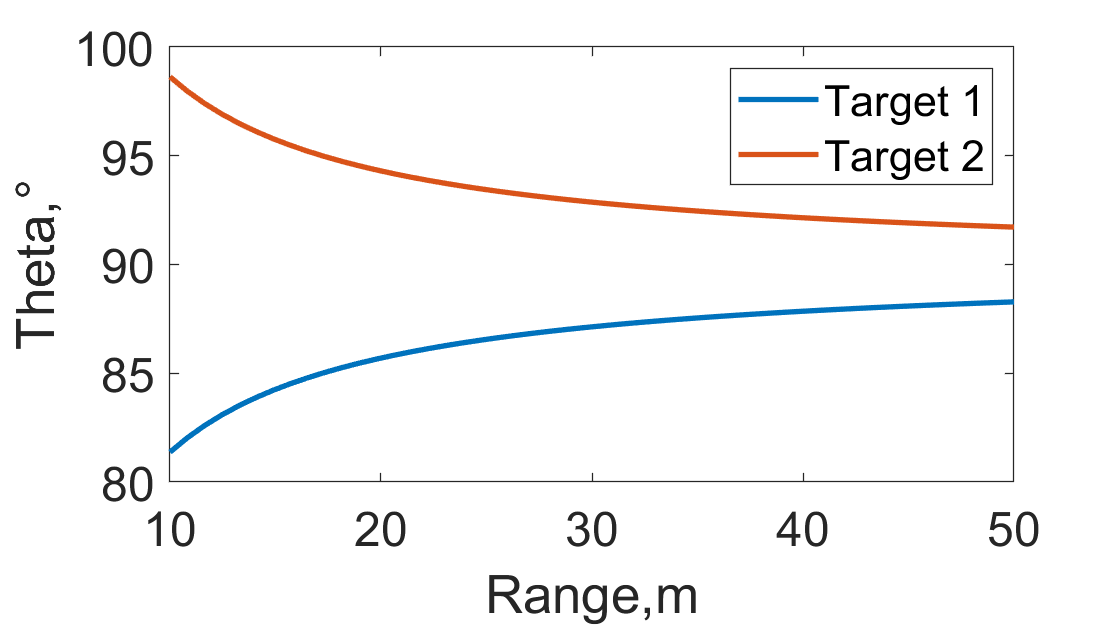}}
  \caption{(a) Simulation scenes. (b) Directions of arrival.}
\end{figure}

Through the above simulation experiments, we prove the advantages of the proposed algorithm FNLANM in DoA estimation. In this section, we simulate an actual driving scenario to visualize it. For this purpose, corner reflectors are used to obtain the point target radar response. As shown in \figref{simuscene}, the radar sensor remains stationary and the two corner reflectors are 3 m apart and moving side-by-side in the direction away from the radar sensor. \figref{1050simu} shows the directions of arrival of the two corner reflectors, as the range increases from 10 m to 50 m. According to the setting of the simulation scene, the angle of the two corner reflectors is continuously changed with the range, and the minimum angular separation of the two corner reflectors is $3.4^{\circ}$.

\begin{figure}[htbp]
  \centering
  \subfloat[]{\label{105030dbf}\includegraphics[width=0.6\linewidth]{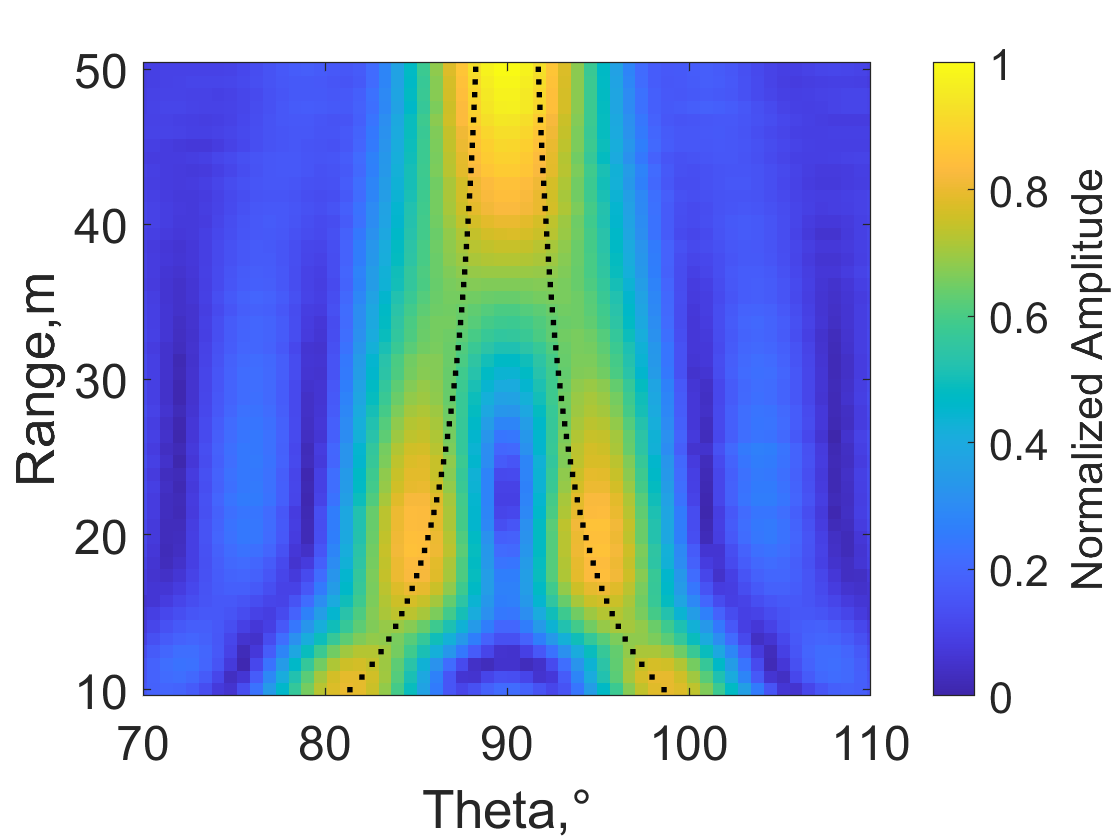}}
  \vfill
  \subfloat[]{\label{105030ist}\includegraphics[width=0.6\linewidth]{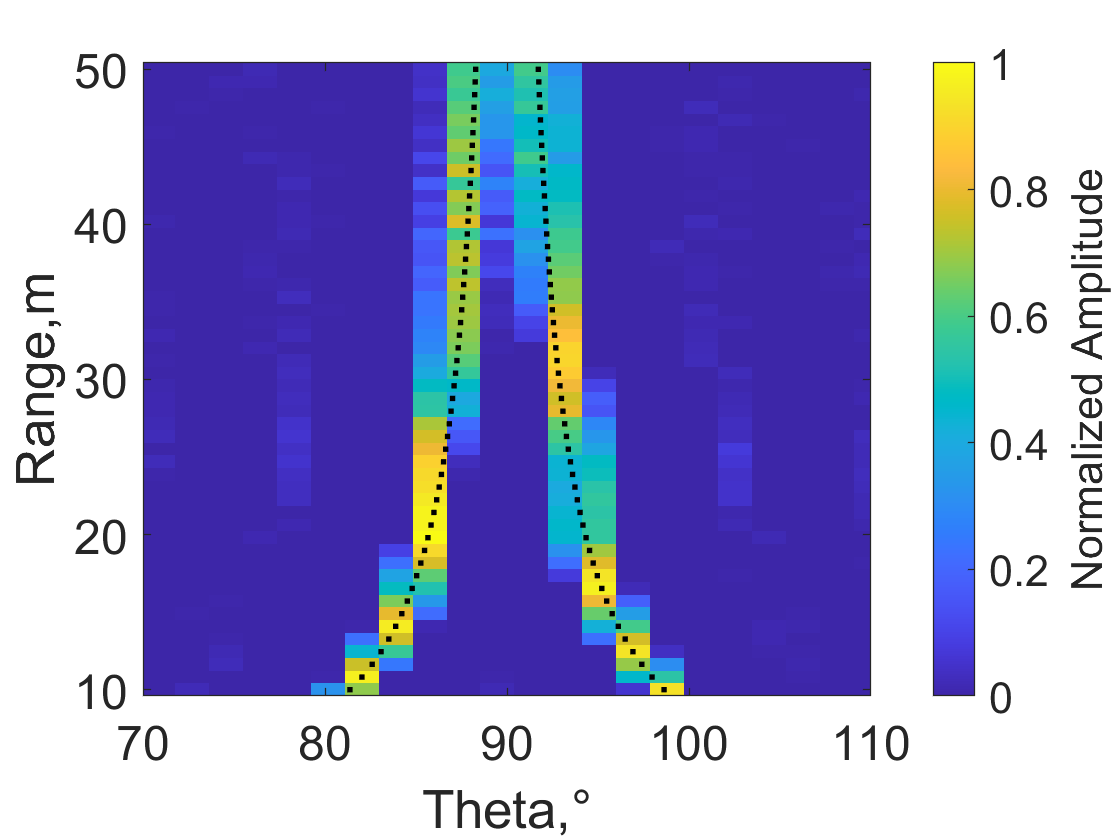}}
  \vfill
  \subfloat[]{\label{105030anm}\includegraphics[width=0.6\linewidth]{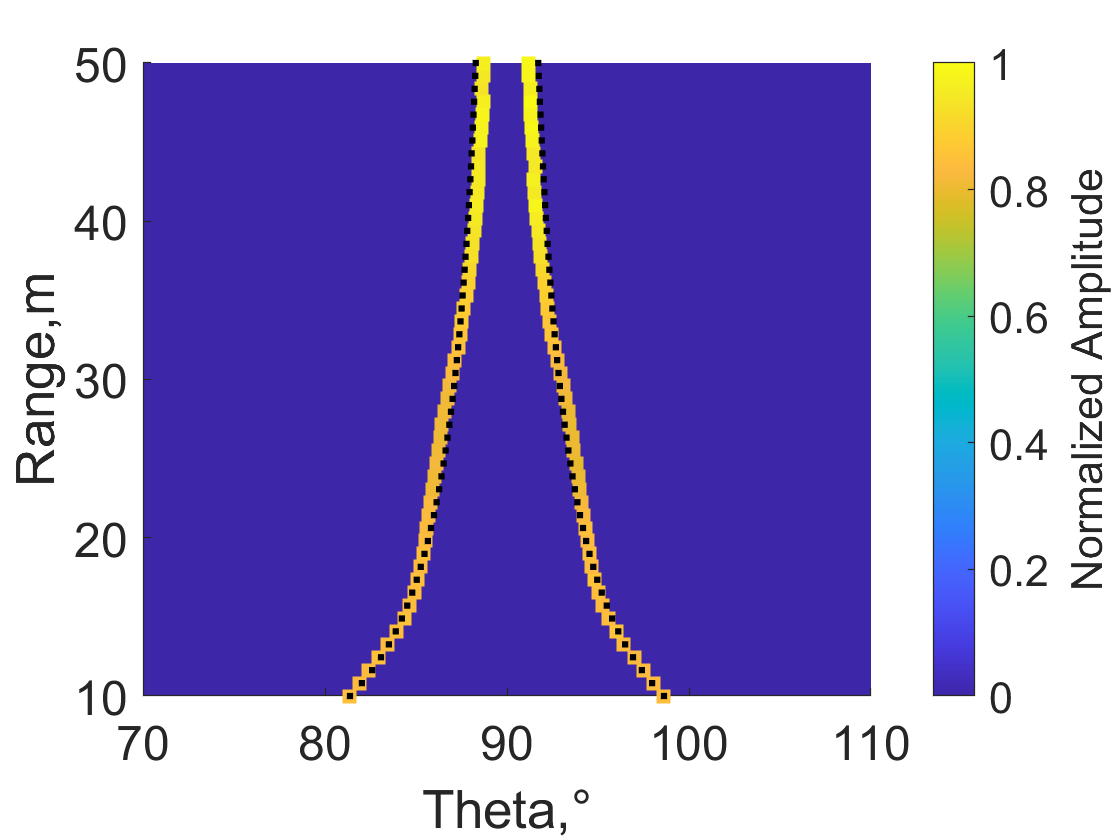}}
  \caption{DoA results of scenario simulation. (a) DBF. (b) Grid-based CS. (c) FNLANM.}
  \label{105030sensimu}
\end{figure}

The simulation parameters are the same as \tabref{taboff} and the results of DBF, grid-based CS and FNLANM for the simulated scenario are shown in \figref{105030sensimu}. We use two black lines to indicate the ground truth of the two corner reflectors.

\figref{105030dbf} shows that DBF is less affected by the off-grid effect, but the resolution is poor. When the angular separation of two corner reflectors is less than the theoretical resolution $\rho$, DBF cannot separate them. As can be seen from \figref{105030ist}, the results of the grid-based CS method are heavily affected by the off-grid effect. More spurious targets appear and the estimated reflectivities of the corner counters have large errors. It is evident from \figref{105030anm} that FNLANM always locates two corner reflectors precisely, with no off-grid effect at all, and can reconstruct both corner reflectors accurately even when the angular separation between two corner reflectors is significantly less than the theoretical resolution.

This part of the experiment visualizes the off-grid effect and the advantages of FNLANM by continuously varying the DoA of two corner reflectors.

\section{Measurement Experiments}\label{sec:real}

\subsection{Measurement setup}
We use two sets of measured data to demonstrate that the FNLANM algorithm proposed in this paper can be directly applied to current radar systems. The two radars used in the measurement experiments are both designed by Beijing Autoroad Tech. Co. Ltd, and the radar parameters are shown in the following \tabref{tab3}. Radar 1 is a 77 GHz short range radar with 600 MHz bandwidth, 3 transmitting and 4 receiving antenna. Radar 2 is a 77 GHz 4D automotive radar with 2 GHz bandwidth, 4 transmitting and 8 receiving antenna.

\begin{table}[htbp]
  \begin{center}
    \caption{Radar parameters.}
    \label{tab3}
    \begin{tabular}{ c  c  c }
                               & \textbf{Radar 1} &                          \\
      \toprule
      Parameters               & Symbols          & Values                   \\
      \midrule
      Frequency,GHz            & $f$              & 77.5                     \\
      Bandwidth,GHz            & $B$              & 0.6                      \\
      Transmitting antenna,mm  & Tx               & 0,5.7,11.4               \\
      Receiving antenna,mm     & Rx               & 0,3.8,7.6,11.4           \\
      Theoretical resolution,° & $\rho$           & 11.9                     \\
      \bottomrule
      \\
                               & \textbf{Radar 2} &                          \\
      \toprule
      Parameters               & Symbols          & Values                   \\
      \midrule
      Frequency,GHz            & $f$              & 77.9                     \\
      Bandwidth,GHz            & $B$              & 2                        \\
      Transmitting antenna,mm  & Tx               & 0,8,38,68                \\
      Receiving antenna,mm     & Rx               & 0,28,32,52,64,78,104,118 \\
      Theoretical resolution,° & $\rho$           & 1.5                      \\
      \bottomrule
    \end{tabular}
  \end{center}
\end{table}

\begin{figure}[htbp]
  \centering
  \includegraphics[width=0.6\linewidth]{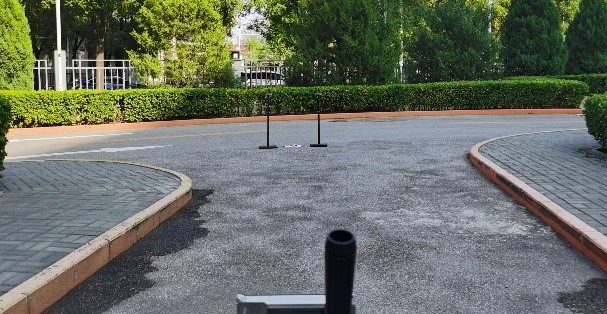}
  \caption{Realistic test scenario.}
  \label{realscene}
\end{figure}

As shown in \figref{realscene}, the data were collected in a park with 1 or 2 corner reflectors as the main targets. The radar calibration matrices were measured under darkroom turntable conditions with an interval of 0.5°. By analyzing the calibration matrices, the precise position of each array element and the channel phase can be obtained, and an accurate observation matrix can be constructed. The observation matrix used in the experiment in this section is obtained by the above method. Moreover, there may be bias in the experimental field measurements or bias in the calibration process. The maximum likelihood (ML) estimate should be considered as the ground truth when validating DoA estimation methods, considering that ML is the optimal estimate under the Gaussian noise model \cite{ml}.

As we can see from the radar parameters, automotive radars are usually based on MIMO radars, and virtual arrays are hardly ULAs, so it makes sense to study DoA estimation algorithms that are applicable to arbitrary array forms. Also, in the real world, targets can be located anywhere in space, so those located on the grids are very few, and most of them are off the grids. Therefore it is meaningful to study gridless sparse DoA estimation applicable to arbitrary arrays.

\subsection{Measurement results}

\begin{figure}[htbp]
  \centering
  \subfloat[]{\label{fu6}\includegraphics[width=0.45\linewidth]{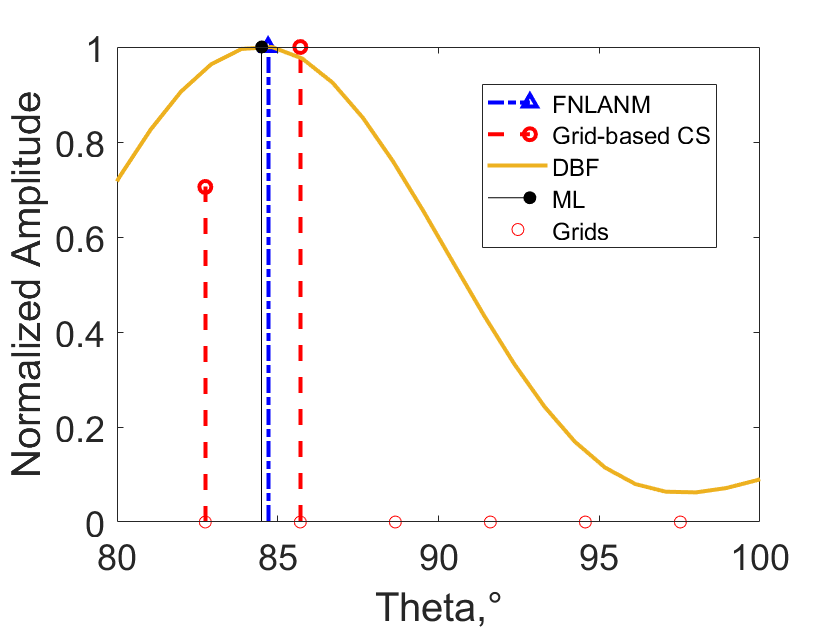}}
  \hfill
  \subfloat[]{\label{6}\includegraphics[width=0.45\linewidth]{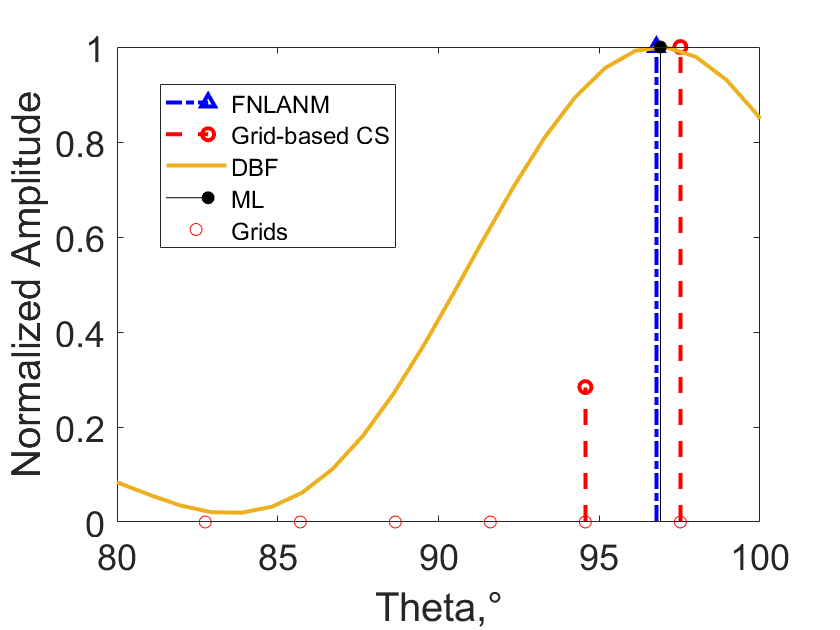}}
  \vfill
  \subfloat[]{\label{12}\includegraphics[width=0.45\linewidth]{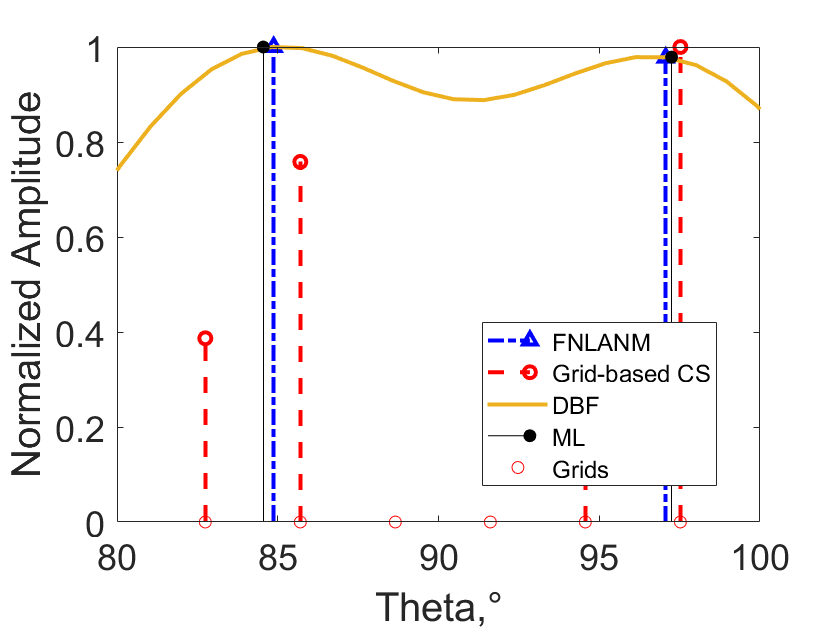}}
  \hfill
  \subfloat[]{\label{fu4}\includegraphics[width=0.45\linewidth]{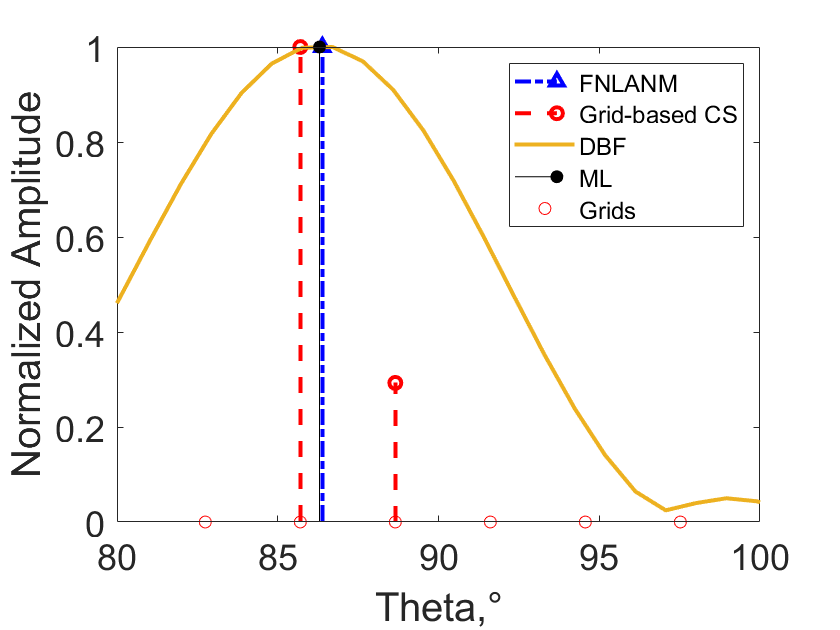}}
  \vfill
  \subfloat[]{\label{4}\includegraphics[width=0.45\linewidth]{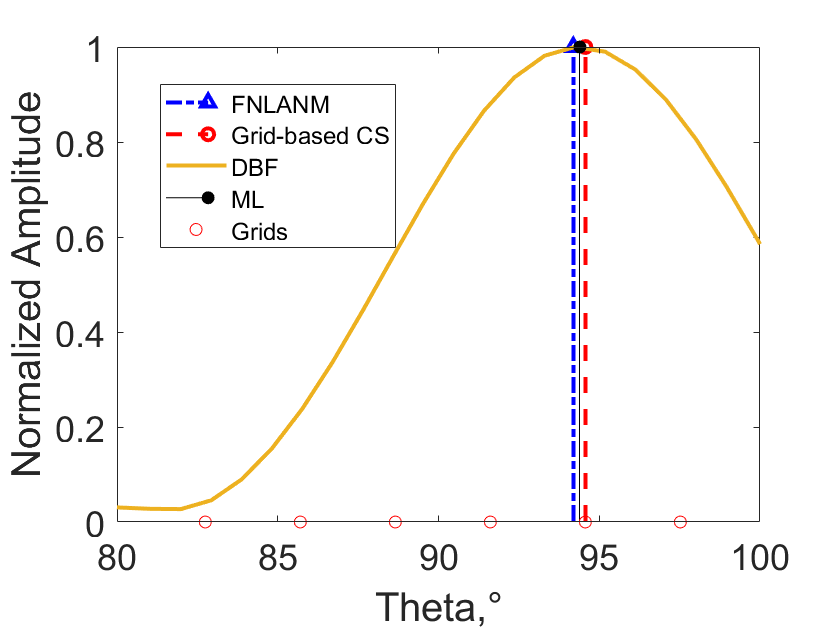}}
  \hfill
  \subfloat[]{\label{8}\includegraphics[width=0.45\linewidth]{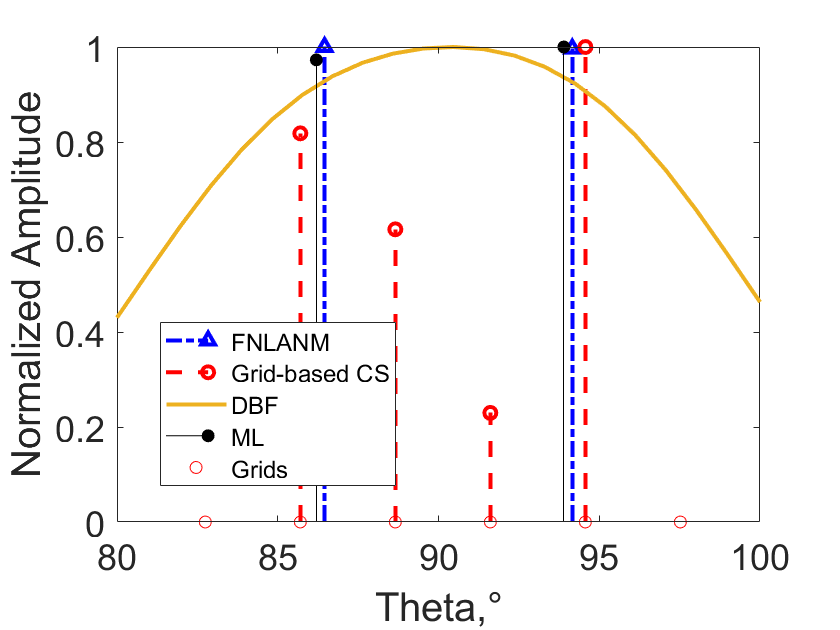}}
  \vfill
  \caption{DoA estimation results of radar 1.}
  \label{real1}
\end{figure}

A total of six experiments are conducted on radar 1, four of which are single corner reflector experiments and two of which are double corner reflectors experiments. The DoA estimation results for the data acquired by radar 1 are shown in \figref{real1}. From the DoA estimation results of the single corner reflector, it can be seen that both DBF and FNLANM can estimate the DoA accurately, while the performance of the grid-based CS method is affected by the off-grid error of the target. \figref{12} shows the results when the corner reflectors in \figref{fu6} and \figref{6} are present simultaneously, and the angular separation of them is $\Delta\theta=12.7^{\circ}$, which is greater than the theoretical resolution. Experimental results show that only the grid-based CS method suffers from the off-grid effect, while both DBF and FNLANM are able to separate and accurately locate the two targets with $\Delta\theta>\rho$. Just as \figref{12}, the targets of \figref{8} are a combination of \figref{fu4} and \figref{4}, with the two corner reflectors separated by $\Delta\theta=7.7^{\circ}$. The experimental results show that FNLANM still exhibits excellent DoA estimation performance when $\Delta\theta<\rho$, which is much better than that of DBF and grid-based CS.

\begin{figure}[htbp]
  \centering
  \subfloat[]{\label{4degdataresult}\includegraphics[width=0.45\linewidth]{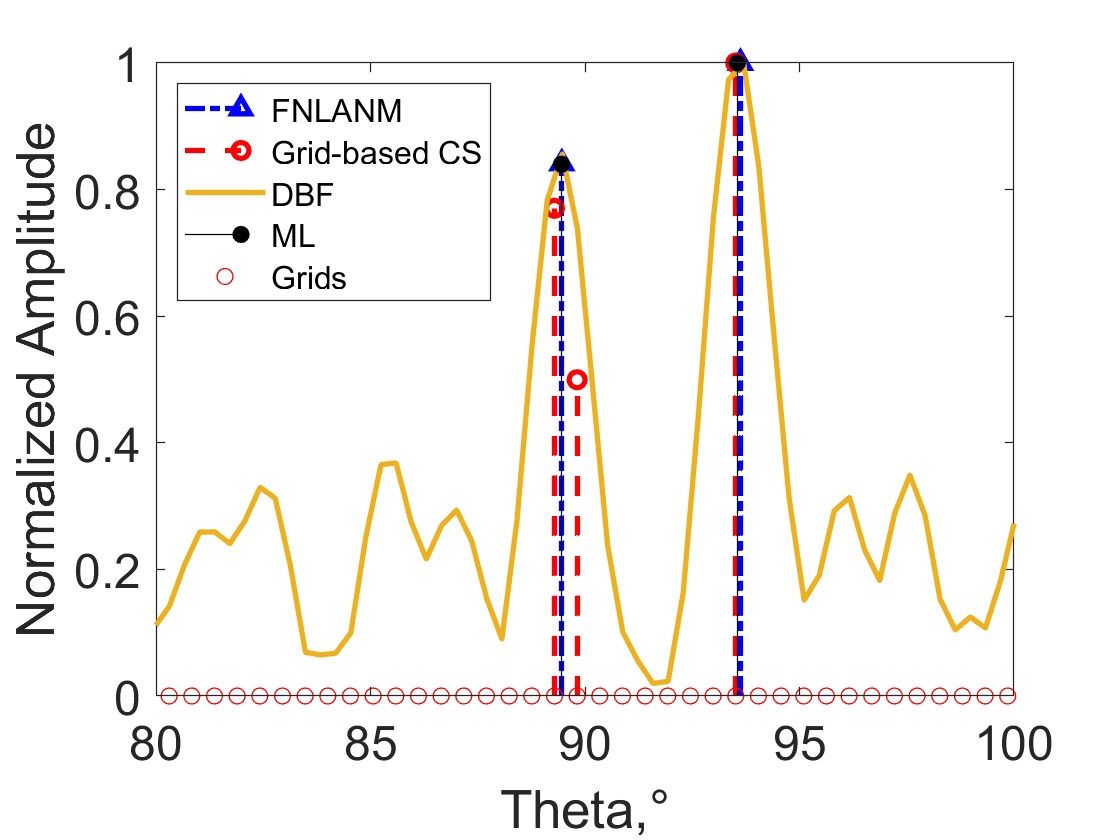}}
  \hfill
  \subfloat[]{\label{3degdataresult}\includegraphics[width=0.45\linewidth]{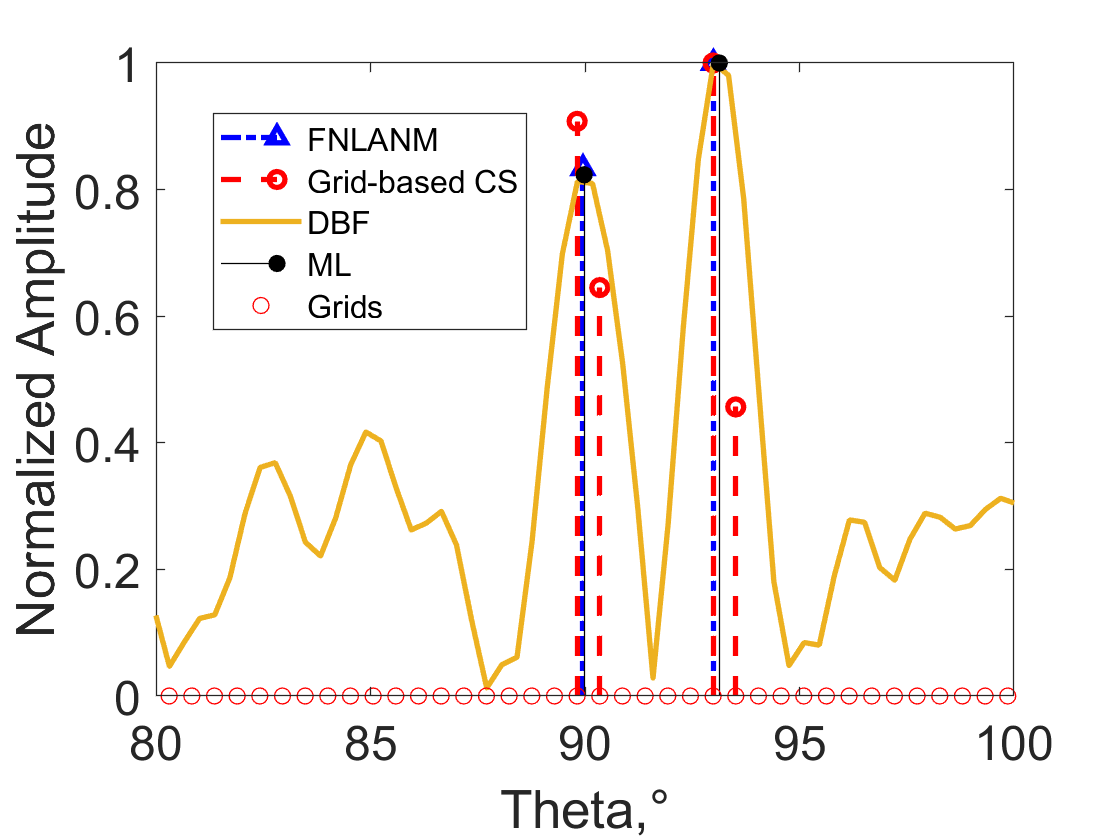}}
  \vfill
  \subfloat[]{\label{2degdataresult}\includegraphics[width=0.45\linewidth]{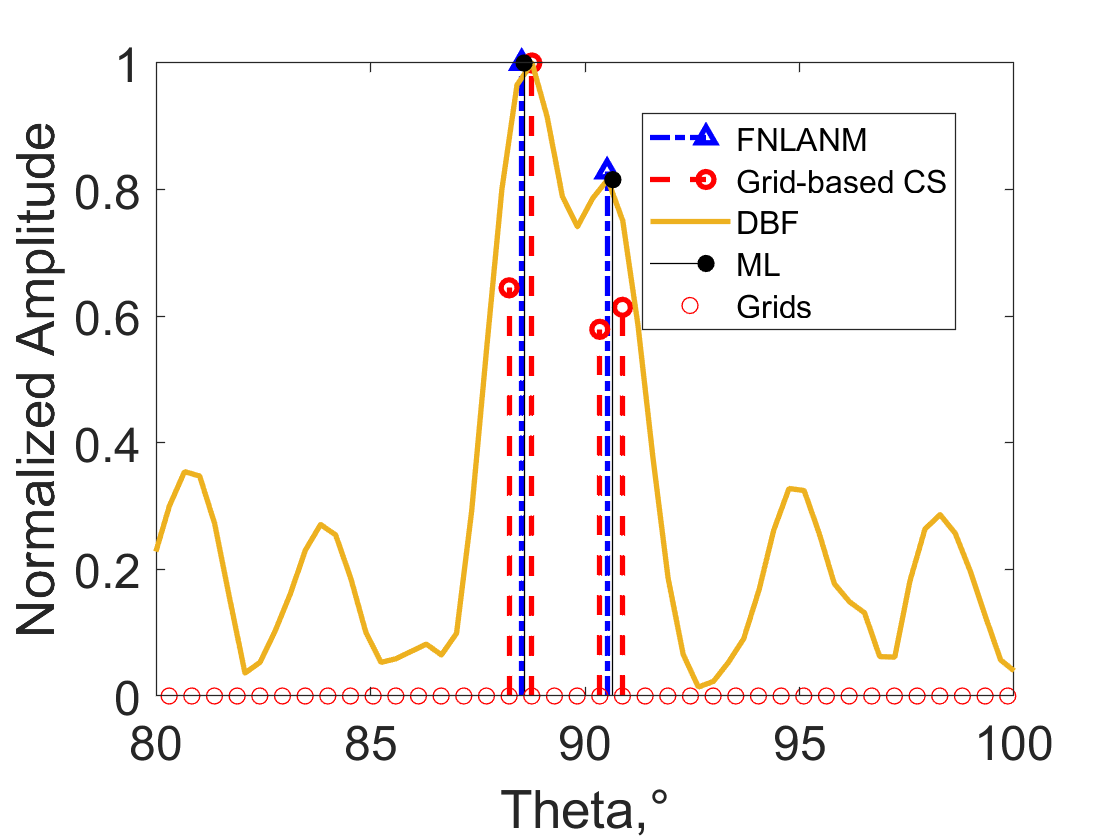}}
  \hfill
  \subfloat[]{\label{1degdataresult}\includegraphics[width=0.45\linewidth]{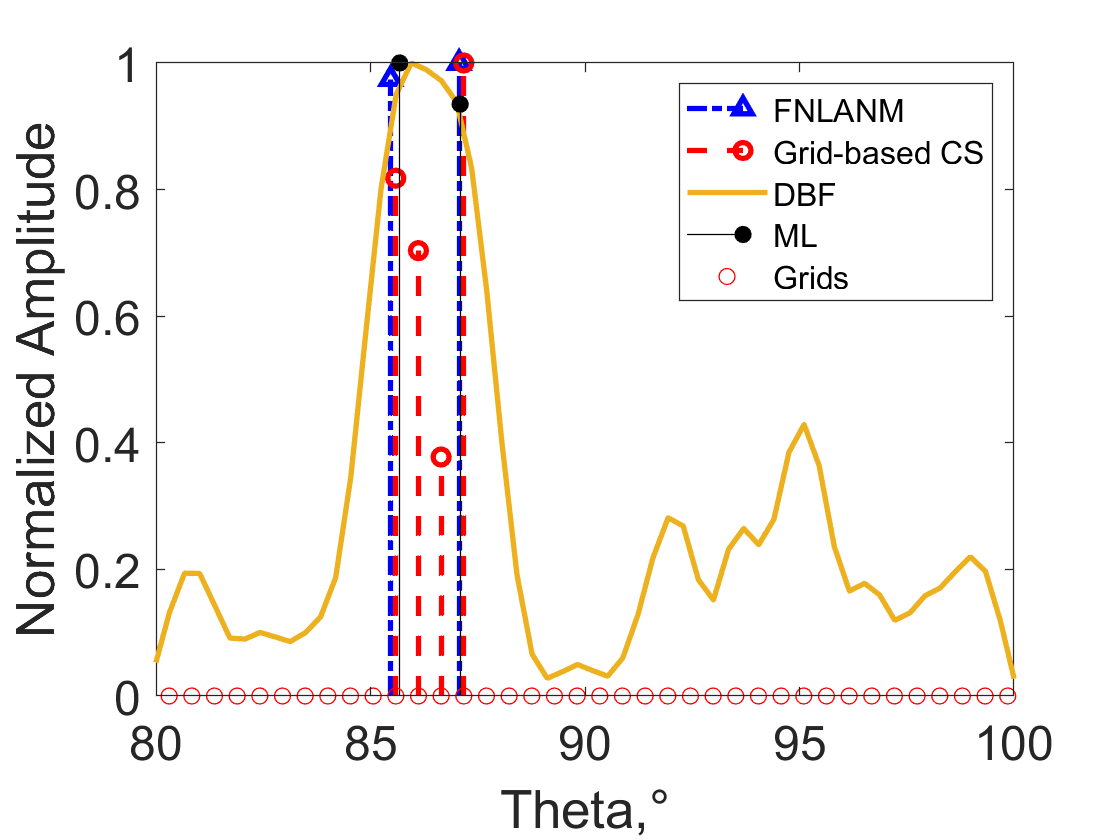}}
  \vfill
  \caption{DoA estimation results of radar 2. (a) $\Delta\theta = 4.2^{\circ}$, (b) $\Delta\theta = 3.1^{\circ}$, (c) $\Delta\theta = 2.1^{\circ}$, (d) $\Delta\theta = 1.4^{\circ}$.}
  \label{real2}
\end{figure}

The theoretical resolution of radar 2 is better than that of radar 1. All the experiments conducted on radar 2 are double corner reflectors experiments, which are mainly used to investigate the resolution capability of the proposed DoA estimation algorithm. The DoA estimation results for the data acquired by radar 2 are shown in \figref{real2}. During the four experiments, the angular separation of the two corner reflectors gradually decreases, and FNLANM is always able to accurately locate their directions and separate them, regardless of whether the targets are off-grid or not.

The experimental results of measured data acquired by automotive MIMO radars from Beijing Autoroad Tech Co., Ltd are compatible with the analysis of theory and simulation experiments. When the target is not precisely located in the grids, there is a high possibility that the energy of the off-grid target will leak to several grids near the ground truth, leading to inaccurate DoA estimates and even false targets of the grid-based CS method. The measurement experiments have verified that FNLANM fundamentally avoids the grid mismatch problem and has super-resolution capability.

\section{Conclusion}\label{sec:con}
In this paper, the gridless DoA estimation problem for NLAs is studied.
The author proposes a fast gridless DoA estimation algorithm for NLAs based on ANM, which combines the array manifold separation technique and a priori knowledge based acceleration technique to reduce the computational complexity by 1.5 orders.
The simulation results indicate that the proposed FNLANM algorithm successfully extends ANM algorithm to arbitrary linear arrays with acceptable computational complexity.
The proposed algorithm is also validated on the automotive MIMO radars designed by Beijing Autoroad Tech Co., Ltd.

Future work will focus on the super-resolution ability and further reduction of the computational complexity, as well as the noise robustness.

\bibliographystyle{IEEEtran}
\bibliography{ref}

\end{document}